\documentclass[trackchanges, twocolumn]{aastex701}

\usepackage{amsmath}
\usepackage{xspace}

\newcommand{\Gaia}{{\it Gaia}\xspace}

\newcommand{\Msun}{\ensuremath{\mathrm{M}_\odot}}
\newcommand{\osim}{\mathord{\sim}} 

\usepackage[acronym]{glossaries}
\makeglossaries
\glsdisablehyper   
\newacronym{MI}{MI}{Mutual Information}

\begin{document}

\title{Galactic Amnesia: The Information Washout of the Milky Way Merger History}

\author[0000-0003-2806-1414]{Lina Necib}
\affiliation{Department of Physics and Kavli Institute for Astrophysics and Space Research, MIT, Cambridge, MA 02139, USA}
\email[show]{lnecib@mit.edu}

\author[0000-0002-1544-1381]{Dylan Folsom}
\affiliation{Department of Physics, Princeton University, Princeton, NJ 08544, USA}
\email{dfolsom@princeton.edu}

\author[0000-0001-5996-4072]{Elliot Y. Davies}
\affiliation{Department of Physics and Kavli Institute for Astrophysics and Space Research, MIT, Cambridge, MA 02139, USA}
\email{eydavies@mit.edu}

\author[0000-0003-3954-3291]{Nathaniel Starkman}
\altaffiliation{Brinson Prize Fellow}
\affiliation{Department of Physics and Kavli Institute for Astrophysics and Space Research, MIT, Cambridge, MA 02139, USA}
\email{starkman@mit.edu}

\author[0009-0005-1250-1800]{Andreas Thoyas} 
\affiliation{Department of Physics, Northeastern University, Boston, MA 02115, USA}
\email{athoyas@mit.edu}




\begin{abstract}

The merger history of a galaxy leaves imprints on its present-day stellar chemodynamics, yet dynamical processes progressively erase this record. We ask: how far back in time, and from which observables, can a galaxy's assembly history still be recovered? We provide a quantitative framework to address this question, using Mutual Information normalized by Shannon entropy to measure how much present-day stellar chemodynamics retains about each past merger's stellar mass $M_\star$ and infall time $t_{\rm infall}$. This framework is applied to TNG50 Milky Way--like galaxies, with comparison to FIRE-2. We find that the gravitational potential and total energy are the most informative and longest-lived tracers of merger properties, highlighting the need for accurately measuring the Milky Way's potential. The information carried by the radial velocity decays to the noise floor within $\sim$5~Gyr, angular momentum carries low information overall with a mass-dependent decay, and chemical abundances retain a flat, low information floor. Information washout depends on three key factors: (1) radial position---stars in the inner galaxy lose information faster due to shorter orbital times; (2) infall time---old mergers are largely phase-mixed; and (3) merger mass---larger mergers sink to the bottom of the potential well via dynamical friction, inducing violent relaxation that erases dynamical information. At each galactocentric radius, we map the observational horizon in the $(M_\star,\; t_{\rm infall})$ plane beyond which past mergers can no longer be recovered from that observable. By recasting merger reconstruction into this quantitative, observable-by-observable map of what is and is not recoverable, our results provide a foundation for interpreting chemodynamical signatures of past mergers and for guiding surveys and modeling toward the observables that maximize merger information recovery.

\end{abstract}



\section{Introduction}\label{sec:intro}

In the $\Lambda$~Cold Dark Matter paradigm, galaxies grow hierarchically: small dark matter halos collapse first and then merge and accrete to form progressively larger systems \citep[see, e.g.,][]{White&Rees_1978,1991ApJ...367...45C,Cooper+:2010:GalacticStellarHaloes}. In this picture, the present-day stellar components of Milky Way--mass galaxies retain fossil records of that assembly in the form of tidal debris, stellar halos, altered disk structure, and redistributed globular cluster systems \citep[see, e.g.,][]{1962ApJ...136..748E,1978ApJ...225..357S,Johnston:1996sb,Johnston:1997fv,1999Natur.402...53H,1999MNRAS.307..495H}. These signatures span a wide range of scales and times: from coherent cold streams produced by recent accretion to  phase-mixed components that reflect early, massive mergers \citep{ Bullock:2005pi,2008A&ARv..15..145H,2020ARA&A..58..205H}. The impact of mergers depends on their mass ratio at accretion, orbital parameters, and timing, all of which determine whether an event contributes to a diffuse stellar halo \citep{Purcell:2007tr,2008MNRAS.391.1806V}, builds a thickened disk~\citep{1993ApJ...403...74Q,1996ApJ...460..121W}, or triggers central star formation~\citep{1991ApJ...370L..65B,1994ApJ...425L..13M,1996ApJ...464..641M,2008MNRAS.384..386C}, and bulge growth \citep{2010ApJ...715..202H}. 

The \Gaia space mission has transformed our ability to read the Milky Way's assembly history by delivering high-precision astrometry (positions, parallaxes, proper motions) and, in later releases, complementary radial velocities and astrophysical parameters for unprecedented numbers of stars~\citep{Gaia:2016zol,gaia_dr3}. With full 6D phase-space information for millions of stars in the Solar neighborhood and increasingly large samples at greater distances, kinematics are no longer a secondary, circumstantial diagnostic but a primary axis along which to separate accreted from in-situ populations and to time and characterize past merger events \citep[see, e.g.,][for a review]{2020ARA&A..58..205H,73685edc5b16025ae1cbf61b5f7661b6cd2f6d06}. Action-angle and integrals-of-motion analyses~\citep{2000MNRAS.319..657H}, clustering in energy-angular-momentum space~\citep{Koppelman:2019,2020ApJ...901...48N,2023MNRAS.521.2623O}, and the identification of coherent streams in phase space~\citep{2018MNRAS.477.4063M,2021ApJ...914..123I,2022MNRAS.509.5992S,2024MNRAS.529.4745S,2025NewAR.10001713B} have all been made feasible or greatly enhanced by \Gaia. This has led to a rapid re-mapping of the Galaxy's fossil record~\citep[see, e.g.,][]{2020MNRAS.498.2472K,2020ApJ...901...48N,Sante2026}, including the recognition of major ancient accretion events; the most dominant latest merger is that of the \Gaia Sausage Enceladus~\citep[GSE,][]{2018MNRAS.478..611B,2018Natur.563...85H}, a galaxy that has merged with the Milky Way approximately 10 billion years ago~\citep{Montalban2021,Naidu2021}, bringing in a chemically and dynamically distinct stellar population~\citep{Lancaster_2018,2019ApJ...874....3N,Feuillet2020,Feuillet_2021_GSE_ages,Iorio2021}  as well as a dark matter component~\citep{2020ApJ...903...25N,Naidu2021}.

Reconstructing the merger history of the Milky Way from its present-day stellar populations is fundamentally a problem of information retrieval under progressive erasure.
Each accretion event deposits stars into the host galaxy with distinctive kinematics, reflecting the orbit and mass of the progenitor at the time of infall.
Over time, however, two physical processes act to erase these signatures.
\textit{Phase mixing} spreads stars from a disrupted satellite along their orbital torus: differences in orbital frequency $\Omega(J)$ cause stars at slightly different actions to drift apart in angle, and although the fine-grained distribution function is conserved, its filaments on the torus become arbitrarily thin so that any coarse-grained observable sees a smooth background in place of the original kinematic substructure.~\citep[e.g.,][]{Johnston:1996sb,Johnston:1997fv,1999MNRAS.307..495H,1999MNRAS.307..877T}.
\textit{Violent relaxation}, operating during and immediately after the merger itself, scrambles the orbital energy distribution of stars as the gravitational potential fluctuates rapidly~\citep{1967MNRAS.136..101L}; a major merger can erase dynamical memory on a single dynamical time, leaving behind a more virialized stellar component with a less coherent kinematic signature.
The efficiency of both processes depends strongly on where and when the merger occurred: inner-halo stars on short orbital periods mix faster than outer-halo stars~\citep[e.g.,][]{Font2005,Johnston2008}, and ancient high-mass mergers are more strongly erased than recent or minor ones~\citep{JeanBaptiste2017,Thomas_unravel_accreted_structures_2025}.
As a result, the dynamical information available to reconstruct a given merger event is not a fixed quantity but degrades with time, and any inference method---whether traditional chemo-dynamical tagging~\citep[see, e.g.,][]{2019ApJ...874....3N,2020ApJ...901...48N} or machine learning~\citep[][]{2020ApJ...903...25N,2020MNRAS.498.2472K,2023MNRAS.521.2623O,Sante2026}---is ultimately limited by how much information survives in the present-day phase space~\citep{Thomas_unravel_accreted_structures_2025}. 

Cosmological hydrodynamical simulations now produce Milky Way--like galaxies with realistic disks, stellar halos, and surviving satellite populations, enabling direct, physically motivated tests of merger-driven assembly scenarios~\citep[][]{Hopkins:2013oba,Schaller:2015vsa,Wetzel2016,2016MNRAS.457..844F,2017MNRAS.467..179G,sanderson2018, 2019MNRAS.490.3196P, 2019ComAC...6....2N, 2021MNRAS.503.5826A, Wetzel2023, 2024MNRAS.532.1814G}. Although building merger histories in simulations is a challenging task, developments in halo finders have improved our ability to track substructure and merger events over time~\citep[e.g.,][]{2001MNRAS.328..726S,Behroozi:2011ju,2013ApJ...763...18B,2013MNRAS.436..150S,2026MNRAS.tmp..279K}. In this work, we focus on TNG50~\citep{2019MNRAS.490.3234N,2019MNRAS.490.3196P}, which combines high resolution (baryon mass $\sim 8.5 \times 10^4~\Msun$) with a large cosmological volume (51.7~Mpc)$^3$ to produce a statistical sample of Milky Way--like galaxies with diverse assembly histories~\citep{2024MNRAS.535.1721P,2025ApJ...983..119F,2025arXiv251021914S}.

In this work, we quantify how much information about a galaxy's merger history 
is encoded in the present-day chemodynamics of its stars, using Milky Way 
analogs from TNG50~\citep{2025ApJ...983..119F,2025arXiv251021914S} and 
merger histories reconstructed from the \textsc{SubFind} and \textsc{SubLink} 
catalogs~\citep{2001MNRAS.328..726S,2009MNRAS.399..497D,2015MNRAS.449...49R}. 
Specifically, we compute the \gls{MI} $\mathcal{I}$ between individual 
chemodynamical observables and the two key properties of each past merger 
event: the progenitor stellar mass $M_\star$ and the lookback time of infall 
$t_\mathrm{inf}$. Using Gaussian processes, we map how $\mathcal{I}(M_\star, t_\mathrm{inf}; r)$ varies as a function 
of Galactocentric radius, merger stellar mass, and elapsed time since accretion.

Applying this framework to GSE-like mergers, we find that radial velocity 
$v_r$ retains essentially no information about the infall time for events older 
than $\osim5$~Gyr. By contrast, the total orbital energy $E$ remains 
informative over longer timescales, while $L_z$ decays substantially with 
time. These results carry direct implications for upcoming spectroscopic 
surveys: \textit{recovering the full merger history of the Milky Way will require 
accurate total energies, and hence precise distances, speeds, and modeling of the gravitational potential, as well as chemical abundances to complement the rapidly decaying kinematic information.}

The paper is organized as follows: In Section~\ref{sec:simulations}, we describe the simulations and the selection of Milky Way--like galaxies. In Section~\ref{sec:features_predictions}, we define the chemodynamical features and merger properties we seek to predict. In Section~\ref{sec:washout}, we employ an information-theoretic framework (\gls{MI}) to quantify how much information about merger histories is encoded in stellar chemodynamics. We examine three key factors controlling information washout: radial position and dynamical timescales (Sec.~\ref{sec:radial_dynamical}), infall time and temporal evolution (Sec.~\ref{sec:temporal_evolution}), and merger mass effects (Sec.~\ref{sec:dominant_mergers}). Section~\ref{sec:implications} discusses implications for the Milky Way, and Section~\ref{sec:conclusions} summarizes our conclusions.

\section{Simulations} \label{sec:simulations}

\subsection{The TNG50 Milky Way--like Sample} \label{sec:tng50}

In this paper, we use 98 Milky Way--like galaxies from the TNG50 simulation~\citep{2019MNRAS.490.3234N,2019MNRAS.490.3196P}, which is part of the IllustrisTNG suite~\citep{2019ComAC...6....2N}. TNG50\footnote{Public releases of these simulations are available at \url{https://www.tng-project.org/data/}.} is a cosmological magnetohydrodynamic simulation, with a (51.7~Mpc)$^3$ volume and gravitational softening lengths 288~pc and $\geq74$~pc for collisionless species and gas, respectively. It is initialized at redshift~127 with 2160$^3$ particles each of dark matter and gas, with dark matter particle mass $4.5\times 10^{5}~\Msun$ and baryon particle mass $\osim8.5\times 10^{4}~\Msun$. Gravity uses the Tree-Particle Mesh algorithm~\citep{2002JApA...23..185B}, while baryonic physics is implemented with the \textsc{Arepo} moving-mesh code~\citep{2010MNRAS.401..791S}, with models for gas radiative processes, star formation, stellar chemical enrichment, supernova-driven galactic outflows, the formation and growth of supermassive black holes, and black hole feedback as described by \citet{2017MNRAS.465.3291W} and \citet{2018MNRAS.473.4077P}.

These Milky Way--like galaxies were selected in~\cite{2025ApJ...983..119F}. We summarize their criteria below:
\begin{itemize}
    \item Stellar mass from $(4-7)\times 10^{10}~\Msun$,
    \item More than 500~kpc from a more massive halo, and
    \item More than 1~Mpc from a halo of mass $10^{13}~\Msun$.
\end{itemize}
No requirement is made on the stellar morphology, but 85\% of the sample is present in the catalog of \citet{2024MNRAS.535.1721P}, which does require a stellar disk. The halos are oriented such that their $z$-axis is parallel to the angular momentum of stars and star-forming gas within twice the half-stellar-mass radius.

To identify the merger history of each galaxy, \cite{2025ApJ...983..119F} used \textsc{SubFind} and \textsc{SubLink} catalogs~\citep{2001MNRAS.328..726S,2009MNRAS.399..497D,2015MNRAS.449...49R} to track the history of each stellar particle to associate it with a certain infalling subhalo. Mergers are taken from the \textsc{SubLink} catalog, and stars are associated to the merger that they spend the greatest number of snapshots bound to.\footnote{We note that this selection was focused on ex-situ stars, as we expect them to be correlated with the merger history of the Galaxy. In-situ stars were not included in the analysis.}

\begin{figure}[t]
    \centering
    \includegraphics[width=0.4\textwidth]{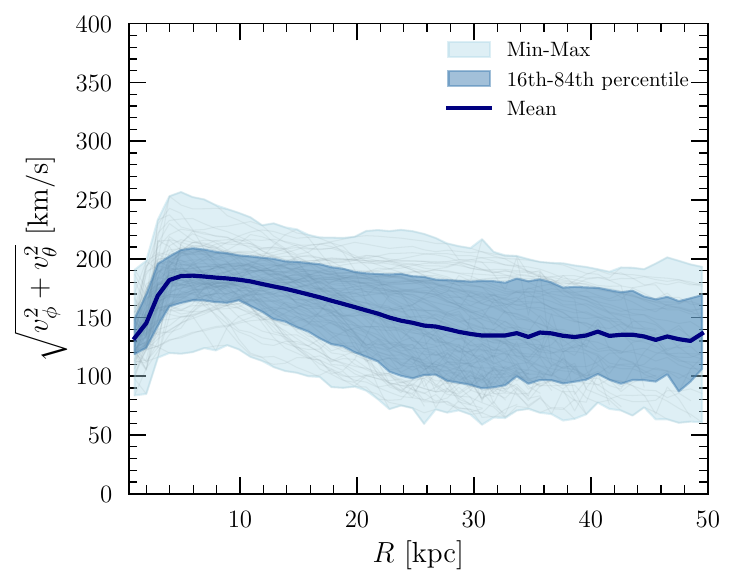}
    \caption{Rotation curves of the 98 simulated galaxies used in this work. The tangential velocities are computed for stars within vertical distances $\leq 5$ kpc of the plane of the galaxy, as a function of the cylindrical radius $R$. The mean in each bin is shown as a solid line, with a dark blue band showing the 16th--84th percentile spread and a pale blue band spanning from the minimum to the maximum. \label{fig:vc}}
\end{figure}

Although all the galaxies that we chose are Milky Way--like, they have slightly different masses, and therefore the features we use are not necessarily on the same scale for all the galaxies. This is evident in Fig.~\ref{fig:vc}, which shows the rotation curves of the 98 galaxies that we consider, computed as the tangential velocity $\sqrt{v_\phi^2 + v_\theta^2}$ of stars within a vertical distance of $\leq 5$ kpc of the disk of the galaxies, as a function of the cylindrical radius of the stars $R$. We find that the mean rotation curve of the galaxies is $\osim 150$ km/s, with a $1\sigma$
 spread of $\sim$100–180 km/s.

Each halo has a total bound mass $M_\mathrm{dyn}$ determined by the \textsc{SubFind} algorithm. Following \citet{2025ApJ...983..119F}, we define a characteristic radius $R_\mathrm{dyn}$ by
\begin{equation}
    M_\mathrm{dyn} = \frac{4}{3}\pi R_\mathrm{dyn}^3 \cdot 200\,\rho_\mathrm{crit},
\end{equation}
for $\rho_\mathrm{crit}$ the critical density of the Universe. Note that $R_\mathrm{dyn} \neq R_{200}$ in general, since $R_\mathrm{dyn}$ is independent of the spatial distribution of the halo; it is purely based on the \textsc{SubFind} mass, which typically includes material from outside $R_{200}$.
For the 98 galaxies we consider, $R_\mathrm{dyn}$ ranges from  [140.0, 318.7] kpc with a median of 220.1 kpc.
We also define a characteristic circular velocity for each galaxy as
\begin{equation} \label{eq:vc_normalization}
v_c = \sqrt{\frac{G \, M_{\rm dyn}(z=0)}{R_{\rm dyn}(z=0)}} \quad [\text{km/s}], 
\end{equation}
with $M_{\rm dyn}(z=0)$ the total dynamical mass of the galaxy at present day, $R_{\rm dyn}(z=0)$ its dynamical radius at present day, and $G$ the gravitational constant. We use $R_\mathrm{dyn}$ and $v_c$ to normalize the features studied in the rest of the study, so that they are on the same scale for all the galaxies.

In this study, we focus on the accreted stars that are within $0.5~R_{\rm{dyn}}$ of the center of the simulated galaxy at redshift $z=0$. This is meant to mimic the reconstruction of the stellar halo by current spectroscopic and astrometric observations~\citep[see, e.g.,][]{2020ApJ...901...48N,Sante2026}.

\subsection{The TNG50 GSE Subsample} \label{sec:gse_subsample}

Additionally, we include criteria to narrow down Milky Way analogs that contain a GSE-like merger by identifying mergers that (i) comprise the majority of their host's ex-situ stellar population within the volume $|z| \in [9,\,15]$~kpc, for $z$ the height out of the disk plane as defined above, and (ii) have radially-biased stellar velocity distributions, with anisotropy $\beta > 0.5$ for its stars in the $|z|\in [9,15]$~kpc volume. This spatial region reflects the volume in the Milky Way within which the GSE is the most significant ex-situ component \citep{2020ApJ...901...48N}, and is in keeping with other studies of GSE analogues in simulated Milky Way--like galaxies~\citep{2019MNRAS.484.4471F, 2025arXiv251021914S}. Of the 98 Milky Way--like galaxies, 32 host such a merger. 

\begin{figure}[t]
    \centering
    \includegraphics[width=0.45\textwidth]{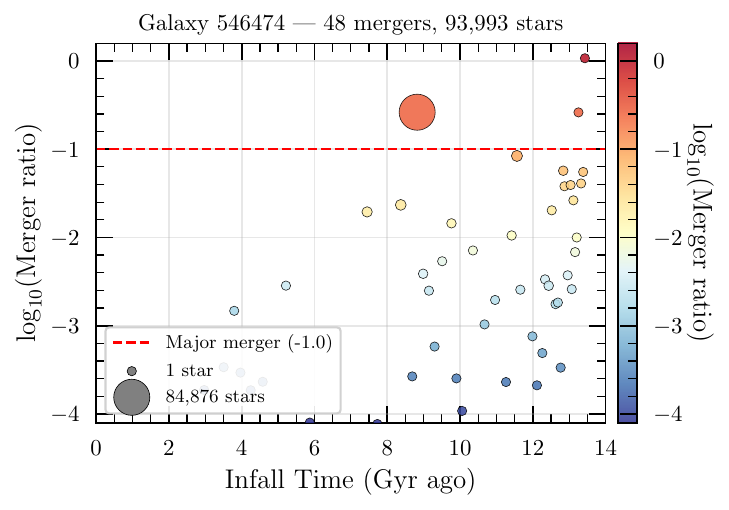}
    \includegraphics[width=0.45\textwidth]{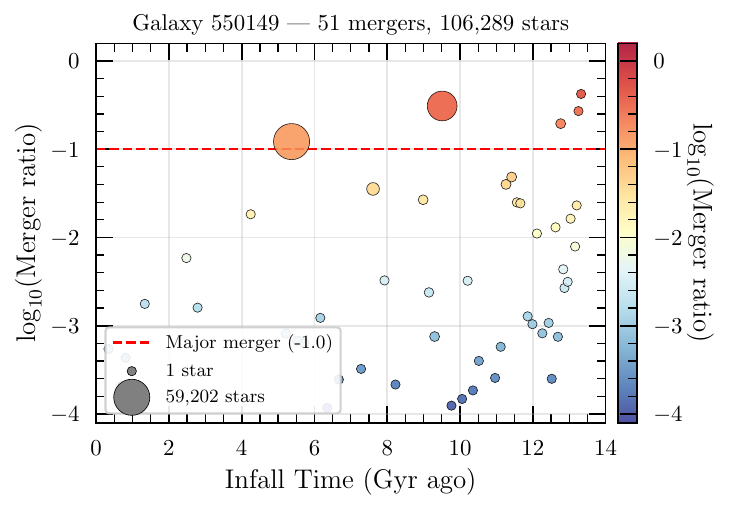}
    \caption{The merger ratio of the different mergers in two distinct galaxies within the GSE-containing sample of~\cite{2025arXiv251021914S}, plotted against the mergers' infall time on the horizontal axis. The size of the circles is linearly interpolated from that of 1 star to the maximal number of stars in the highest merger for each galaxy. The top panel shows a galaxy for which the GSE-like merger is considered dominant, which we include in our own GSE-containing subsample. The galaxy in the bottom panel, however, has a GSE-like merger which is not considered dominant: there is a larger-ratio merger that also contributes many stars, which might complicate the interpretation and the comparison with the Milky Way. For this reason, this galaxy is not included in our GSE-containing subsample. \label{fig:gse_properties}}
\end{figure}

Given our interest in how large mergers affect their hosts, we further restrict our attention to the galaxies that have a single dominant merger that meets the above criteria, and find 19 such galaxies. Here, the dominant merger is defined as the merger with the largest merger-to-host mass ratio among the top five mergers that contributed the most stars within $0.5R_\mathrm{dyn}$, with a minimum of 1000 such stars required to be considered a merger. An example of this selection can be seen in Fig.~\ref{fig:gse_properties}: the top panel illustrates a galaxy with one dominant GSE-like merger, while the bottom panel shows a galaxy with two mergers that contribute a large number of stars, which we exclude from this subsample, as the combined effect of the two mergers can be hard to disentangle. 
We refer to this subsample as the ``GSE-like'' sample, as these mergers are the most similar to the GSE in terms of their contribution to the ex-situ population, their kinematics, and the otherwise quiet merger history of the host galaxy.

The properties of the dominant merger in each of the 19 GSE-like galaxies are summarized in Table~\ref{tab:gse_properties}.  The 19 galaxies in this subsample correspond to SubhaloIDs 613192, 571633, 559386, 552581, 552414, 546474, 543729, 541218, 540920, 537941, 537488, 532301, 530330, 528322, 505100, 504559, 499704, 459558, and 419618. The GSE-like dominant mergers span a wide range of infall times (4--13~Gyr ago), stellar masses ($\log_{10}(M_\star/\Msun) \sim 8$--10), and merger ratios ($\osim1$:15 to nearly 1:1\footnote{There is a single outlier in which the merger was more massive than the host (see footnote~\ref{fn:sublink}).}, with a median of $\osim1$:5). A distribution of the infall times is shown in Fig.~\ref{fig:timing_dominant_merger}, showing two clusters, one at 7--10 Gyr ago and one at 3--6 Gyr ago.

\begin{table}[t]
    \centering
    \begin{tabular}{lcccc}
        \hline
        \hline
        Property & Min & Max & Median & Mean \\
        \hline
        Infall time [Gyr ago] & 3.97 & 12.69 & 9.51 & 9.09 \\
        $\log_{10}(M_{\rm{peak}}/\Msun)$ & 9.50 & 11.62 & 10.91 & 10.82 \\
        $\log_{10}(M_\star/\Msun)$ & 7.85 & 10.27 & 9.03 & 9.11 \\
        $\log_{10}(\mathrm{merger\ ratio})$ & $-1.11$ & $+0.57$ & $-0.61$ & $-0.53$ \\
        \hline
    \end{tabular}
    \caption{Properties of the dominant merger in the 19 GSE-like galaxies. The dominant merger is defined as the merger with the largest merger-to-host mass ratio among the top five mergers contributing the most stars, requiring a minimum of 1000 stars. The distribution of infall times can be found in Fig.~\ref{fig:timing_dominant_merger}. \label{tab:gse_properties}}
\end{table}

\subsection{FIRE-2 Comparison Sample} \label{sec:fire_comparison}

Our conclusions should not depend on the choice of cosmological simulation. We therefore repeat the analysis on an independent sample of Milky Way--like galaxies from the Feedback In Realistic Environments (FIRE) project, which differs from IllustrisTNG in both its hydrodynamics solver and its subgrid physics~\citep{2015MNRAS.450...53H,2017arXiv170206148H,Wetzel2016}. The FIRE-2 simulations\footnote{Public releases of these simulations are available at \url{https://flathub.flatironinstitute.org/fire}~\citep{Wetzel2023,2025arXiv250806608W}} include explicit models for stellar feedback and resolve the multi-phase interstellar medium, which can lead to different merger histories and chemodynamical signatures compared to IllustrisTNG. By comparing results across these two independent simulation suites, we can assess the robustness of our conclusions regarding the information content of ex-situ stars and the effects of mergers on stellar chemodynamics. Furthermore, the FIRE-2 suite comprises zoom-in simulations of relatively isolated galaxies, while the $(51.7~\mathrm{Mpc})^3$ cosmological volume of the TNG50 simulation contains two Virgo mass clusters ($M_\mathrm{dyn} \gtrsim 10^{14}~\Msun$), and $\osim80\%$ of the TNG50 Milky Way--like galaxies are within 16~Mpc of one of them---closer than the actual Milky Way is to the Virgo cluster. The differences in the local environments between simulation suites may also induce differences in the assembly histories, making the cross-check with FIRE-2 even more valuable. 

We focus this analysis on three Milky Way--like galaxies from the FIRE-2 suite: m12i, m12f, and m12m~\citep{Wetzel2016,2017MNRAS.471.1709G,2017arXiv170206148H}. These galaxies have stellar masses of $\osim 10^{10}~\Msun$ at $z=0$ and have experienced a range of merger histories, including both major and minor mergers~\citep[see][for the detailed merger histories of these galaxies]{2026arXiv260325783Z}. Details of their properties are summarized in Table~\ref{tab:fire_properties}. 

\begin{table}[t]
    \centering
    \begin{tabular}{lcc}
        \hline
        \hline
        Simulation & $R_{\rm{vir}}$ [kpc] & $M_{\rm{vir}}$ [$10^{12}\Msun$] \\
        \hline
        m12i & 268.0 & 1.07 \\
        m12f & 293.2 & 1.40 \\
        m12m & 197.8 & 1.47 \\
        \hline
    \end{tabular}
    \caption{Properties of the three FIRE-2 Milky Way--like galaxies: m12i, m12f, and m12m from~\cite{sanderson2018}.\label{tab:fire_properties}}
\end{table}

\section{Information Content of Ex-Situ Stars} \label{sec:features_predictions}

\begin{figure*}[t]
    \centering
    \includegraphics[width=0.8\textwidth,clip,trim=0 0 0 2cm]{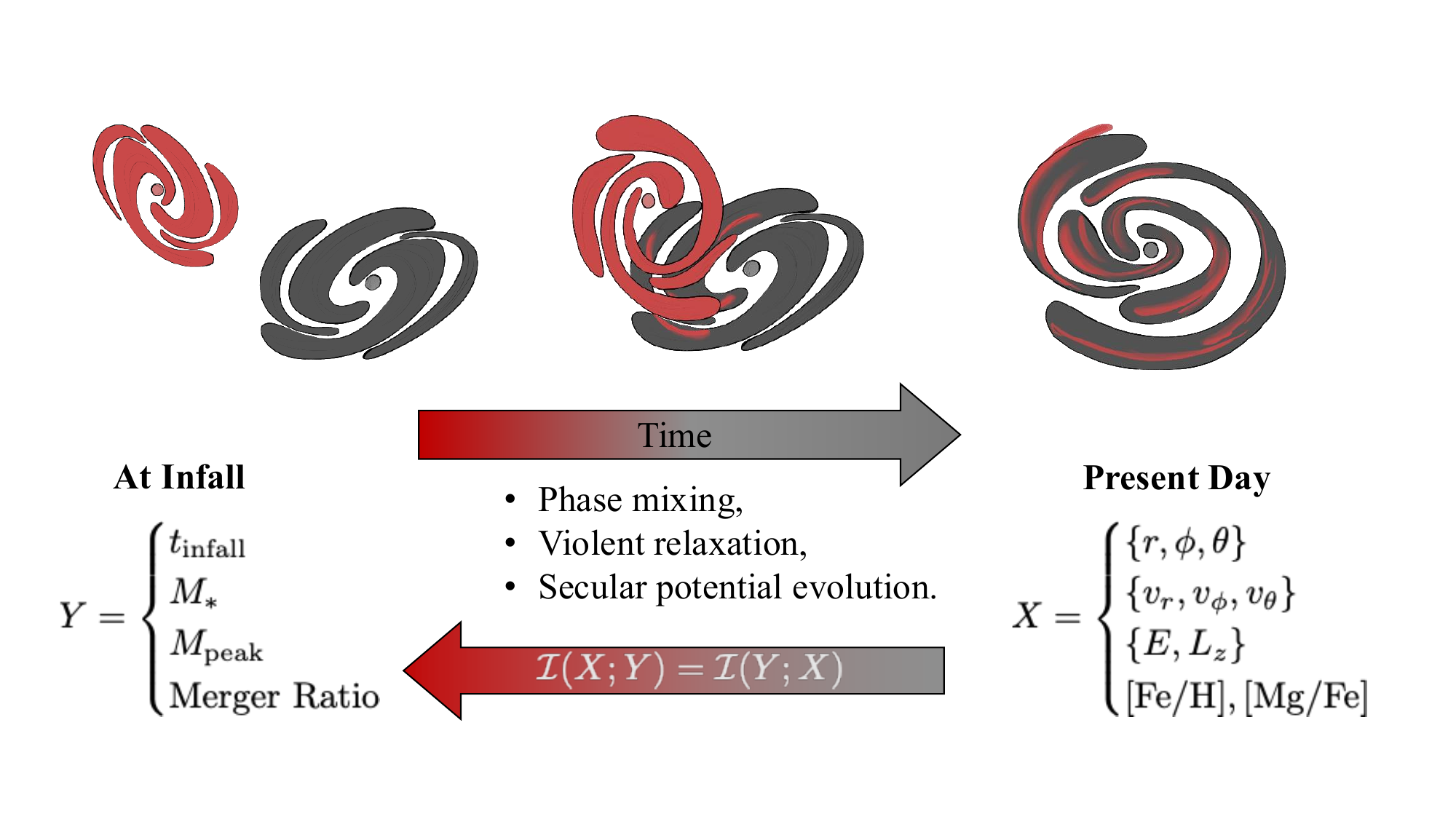}
    \caption{Sketch of the features and predictions discussed in Sec.~\ref{sec:features_predictions}. Features are labeled as $X$, and evaluated at present day, while predictions are labeled as $Y$, and are properties of the merger event at infall time. We evaluate the mutual information $\mathcal{I}(X; Y)$ between the features and the predictions, which quantifies how much information about the merger properties is encoded in the present-day chemodynamics of accreted stars. 
     \label{fig:feature_distributions}}
\end{figure*}

The critical question we seek to answer in this work is how much of the information about a merger (e.g. its stellar mass or infall time) is encoded today in the chemodynamics of accreted stars. To this end, we define a particular set of ``features," variables that may retain information about ``predictions" within specific merging scenarios. We then discuss the amount of information these features hold. A sketch of the features and predictions that we discuss in this section is shown in Fig.~\ref{fig:feature_distributions}\footnote{Artwork by Yasmine Necib.}.

\subsection{Features: Dynamics \& Chemistry} \label{sec:features}

We mainly focus on properties of stars that are accessible to observations. To that end, we use:
\begin{itemize}
    \item \textbf{Positions:} We use galactocentric spherical coordinates for each star: $r$, the radial distance from the center, $\theta$, the polar angle, and $\phi$, the azimuthal angle. All these values are taken at the present day (redshift $z=0$).
    \item \textbf{Velocities:} Similarly, we use the velocities in galactocentric spherical coordinates at present day: $\{v_r, v_\theta, v_\phi\}$.
    \item \textbf{Constants of Motion:} Although the gravitational potential 
    for a specific star is not easily accessible in observations, we do have 
    access to some modeling of the gravitational potential of the Milky 
    Way~\citep[see, e.g.,][]{2015ApJS..216...29B,2017MNRAS.465...76M}. 
    In a static axisymmetric potential, the true constants of motion are the 
    three action variables $(J_R, L_z, J_z)$~\citep{2012MNRAS.426.1324B}, where $J_R$ 
    is the radial action quantifying the amplitude of radial oscillations, 
    $L_z \equiv J_\phi$ is the azimuthal action (the $z$-component of angular 
    momentum, exactly conserved in axisymmetry), and $J_z$ is the vertical 
    action quantifying oscillations perpendicular to the disk plane. In 
    practice, the total orbital energy $E$ and $L_z$ serve as more 
    observationally accessible proxies for this action space, though $E$ 
    is only approximately conserved in a realistic time-varying 
    potential~\citep{BinneyTremaine:2008}. Previous 
    literature~\citep[see, e.g.,][]{2000MNRAS.319..657H, Koppelman:2019, Massari:2019, 2020ApJ...901...48N, Lovdal:2022,2023MNRAS.521.2623O} has exploited clustering in $(E, L_z)$ and 
    $(J_R, L_z, J_z)$ space to identify merger debris and substructure within 
    the Milky Way. In this work, we will only use $E$, $L_z$, and $\Phi$, with $E = \frac{1}{2}v^2 + \Phi$ 
    and $\Phi$ the gravitational potential at the star's present-day 
    position, extracted directly from the simulation. The angular momentum 
    along the $z$-direction is calculated as $L_z = x\,v_y - y\,v_x$, with 
    $x, y$ the Cartesian coordinates of the star within the plane of the disk, 
    and $v_x, v_y$ the corresponding Cartesian velocity components. 
    \item \textbf{Chemical Abundances:} The IllustrisTNG simulations directly track multiple elemental abundances for each star particle. We focus on [Fe/H] and [Mg/Fe]; [Fe/H] is the most commonly available measurement in spectroscopic surveys~\citep[see, e.g.,][]{abdurrouf22}, and has been shown to correlate with progenitor mass via the mass--metallicity relation~\citep{Kirby:2013wna} in previous works~\citep[see, e.g.,][]{Mackereth2018,2019ApJ...874....3N}, and [Mg/Fe] is taken as a proxy for the alpha elements \citep{1979ApJ...229.1046T,1997ARA&A..35..503M} as it has been used
to chemically separate accreted from in-situ populations
in the Milky Way~\citep[see, e.g.,][]{2015MNRAS.453..758H,Mackereth2019}. To obtain both ratios, we use the \texttt{GFM\_Metallicity} field, compute the iron and magnesium mass fractions assuming Solar abundance ratios from~\cite{2009MNRAS.399..574W}, and convert to the logarithmic [Fe/H] and [Mg/Fe] scale relative to Solar. There are limitations to the chemical enrichment model in IllustrisTNG, because it tends to overestimate both [Fe/H] and [Mg/Fe] compared to the Milky Way, as discussed in~\cite{2018MNRAS.477.1206N}. As a point of comparison albeit with much smaller statistics, we use the FIRE-2 simulations, which have a more detailed chemical enrichment model that tracks 11 individual elements, including iron and magnesium~\citep{2017arXiv170206148H}.
\end{itemize}

As mentioned in Sec.~\ref{sec:tng50}, the galaxies in our sample have different masses, and therefore the features we use are not necessarily on the same scale for all the galaxies. To account for that, we normalize each of the features by the relevant scales of the galaxy: its virial radius and its characteristic circular velocity. 
The normalized features are then defined as: 
\begin{align} 
\Phi_{\rm norm} &= \Phi/v_c^2, \quad E_{\rm norm} = E/v_c^2 \nonumber\\
r_{\rm norm} &= r/R_{\rm dyn} \nonumber\\
v_{r,\rm norm} &= v_r/v_c, \quad v_{\theta,\rm norm} = v_\theta/v_c, \quad v_{\phi,\rm norm} = v_\phi/v_c \nonumber\\
L_{\rm norm} &= L/(R_{\rm dyn} \, v_c), \quad L_{z,\rm norm} = L_z/(R_{\rm dyn} \, v_c) \label{eq:normalization}
\end{align}
The remainder of this paper adopts the normalized features, even when the subscript is dropped for simplicity. This normalization is not meant to reduce the scatter of the values, but rather to provide an interpretable set of features with which we can compare galaxies of different scales. 

\subsection{Predictions}\label{sec:predictions}

Using the stellar features established in Sec.~\ref{sec:features}, we seek to study the amount of information held by an accreted star about its progenitor galaxy. More specifically, we aim to determine the following properties of the merger:
\begin{itemize}
    \item \textbf{Infall Time:} This is the lookback time at which a star is accreted onto the main halo, specifically the snapshot just before it is found within the main halo's virial radius.\footnote{For some mergers, the \textsc{SubLink} algorithm is not able to reconstruct the trajectory of the merger and the infalling subhalo is never seen to cross the host's virial radius. In these cases, we take the present day to be the infall time, or disregard the merger if there is no \textsc{SubLink} halo at the present day.}  In cases where the progenitor exits and re-enters the virial radius, the most recent infall is used.
    \item \textbf{Stellar Mass:} This is calculated as the stellar mass of the progenitor at infall, with infall time defined as above. 
    \item \textbf{Peak Mass:} This is the total mass of the progenitor at infall, calculated similarly to the stellar mass above. 
    \item \textbf{Merger Ratio:} The merger ratio is calculated as the total mass of the progenitor divided by the total mass of the main halo at infall.\footnote{\label{fn:sublink}In the \textsc{SubLink} algorithm, the primary halo is the halo with the greatest time-integrated mass history, rather than the halo which instantaneously has the largest mass. This can in some occasions lead to mass ratios greater than one, for which an infalling halo has larger mass than the central host.} Although this information is correlated with the predictions above, we include it given that it incorporates the mass evolution of the main halo as well, taking into consideration the overall assembly history of the simulated galaxy.
\end{itemize}

Note that for the stellar mass, peak mass, and merger ratio, we consider the base-ten logarithm of these quantities. 
Further, we note that a merger's stellar mass, peak mass, and merger ratio are all strongly correlated: larger merger ratios correspond to larger peak masses, and those halos host larger galaxies. Therefore, any feature that recovers information on one of these three labels will also generally recover a similar amount of information on the others, and as such we primarily focus on stellar mass as being representative of each of these three predictions. 

\subsection{Mutual Information} \label{sec:mutual_information}

In order to understand the information content present in the features and the degradation of this information with time, we use the \gls{MI} between each of the features (defined in Sec.~\ref{sec:features}) and the predictions (defined in Sec.~\ref{sec:predictions}). \gls{MI} $\mathcal{I}$ quantifies the amount of information obtained about one random variable $Y$ through observing another random variable $X$ (see Fig.~\ref{fig:feature_distributions}). It is defined as 
\begin{equation}
\mathcal{I}(X;Y) = \sum_{x \in X} \sum_{y \in Y} p(x,y) \log \left(\frac{p(x,y)}{p(x)p(y)} \right)
\end{equation}
where $p(x,y)$ is the joint probability distribution of  $X$ and $Y$, and $p(x)$ and $p(y)$ are the marginal probability distributions of $X$ and $Y$, taken to be the observed features and prediction variables, respectively~\citep{Shannon1948,Information_Theory}. \gls{MI}, which is symmetric in $X$ and $Y$, can be thought of as the reduction in uncertainty about one variable given knowledge of the other, with a value of zero indicating independence of the two variables~\citep{Information_Theory}.

The value of the \gls{MI} is hard to interpret directly, as its scale depends on the ranges of the two variables. Therefore, we normalize the \gls{MI} by the entropy of the prediction variable $H(Y)$~\citep{Shannon1948}, defined as
\begin{equation}
H(Y) = - \sum_{y \in Y} p(y) \log (p(y)).
\end{equation}
This normalization ensures that the \gls{MI} ranges from 0 to 1, where 0 indicates no information shared between the variables, and 1 indicates that knowing one variable completely determines the other. This normalized \gls{MI} provides a clearer measure of association strength between features and predictions in our analysis.

More technically, given that the list of labels are discrete (i.e. the merger properties for stars from the same merger are exactly the same), while the features (stellar energies, positions, etc.) are continuous, we must bin the features to compute the \gls{MI}. To that end, we discretize the features $X$ into bins so that we can compute discrete \gls{MI} similarly to the predictions $Y$. In this work, we compute $H(Y)$ using the \texttt{entropy} function from \texttt{scipy.stats}~\citep{2020SciPy-NMeth} after computing the number of unique mergers and their relative frequencies. For the \gls{MI} $\mathcal{I}(X;Y)$, we choose the bin count as the ${\rm max} (\sqrt{N}, 10)$, where $N$ is the number of stars in the bin considered, and discretize $X$ in equal-occupancy bins. We then compute the \gls{MI} using the \texttt{mutual\_info\_score} function from \texttt{sklearn}~\citep{pedregosa2018scikitlearnmachinelearningpython}, and normalize it by the entropy of the prediction variable $H(Y)$ computed as described above.

\gls{MI} requires variation in both the observable $X$ and 
the label $Y$ to be meaningful. Since all stars from the same merger 
share identical labels by construction, computing \gls{MI} for stars from
a single merger would yield no information. We therefore compute \gls{MI} 
across populations of stars in two complementary ways, depending on the 
quantity of interest. In the first case, we compute \gls{MI} for each galaxy 
individually, using all of its stars: the label distribution $Y$ 
reflects the full set of mergers that contributed stars to that galaxy, 
and the feature distribution $X$ reflects the present-day observables 
of all those stars. In the second case, we compute \gls{MI} for subsets of 
stars defined by a property of interest, such as galactocentric radius 
or progenitor infall time. Here, we pool stars across all galaxies 
within a given bin, so that the sample contains contributions from many 
distinct merger histories and the necessary variation in $Y$ is 
preserved. To avoid adding noise to the system, we only consider the mergers that have a minimum of 50 stars, and only compute the \gls{MI} in bins that include at least 1000 stars and two mergers. In both computations, whether it is for individual galaxies or pooled bins, the entropy $H(Y)$ used for normalization is 
computed from the label distribution of the stars entering that 
particular calculation so that the normalization reflects the actual diversity of mergers 
represented in each sample.

\section{Information Washout Across Galaxy Properties} \label{sec:washout}

The present-day chemodynamical properties of accreted stars encode a fossil record of their host galaxy's merger history. However, this information has a finite lifetime; secular evolution, external gravitational perturbations, orbital phase mixing, and violent relaxation progressively erase the dynamical signatures of past accretion events. In this section, we quantify how much information about merger properties—infall time and stellar mass—is retained in present-day stellar features, and we identify the physical processes that control information washout.

We begin by identifying which stellar features contain the most information about merger properties (Sec.~\ref{sec:importance_chemodynamics}). We find that energy and to a lesser extent angular momentum are the most informative dynamical tracers, while metallicity and [Mg/Fe] provide robust chemical signatures, albeit much lower than the information encoded in dynamics. Velocities seem to be the most susceptible to information washout. However, the amount of information encoded in these features is not uniform across the galaxy; it depends critically on the merger properties, such as the time since infall and the mass of the progenitor, as well as the present-day location of the stars.
In Sec.~\ref{sec:galaxy_props}, we systematically examine how three key factors control information retention. Finally, we will use Gaussian process regression to predict the \gls{MI} as a function of these three factors (Sec.~\ref{sec:functional_form}), which we later use to study the implications for the Milky Way's merger history (Sec.~\ref{sec:implications}).

\subsection{Importance of Chemodynamics in Merger History} \label{sec:importance_chemodynamics}

\begin{figure}[t]
    \centering
    \includegraphics[width=0.9\linewidth]{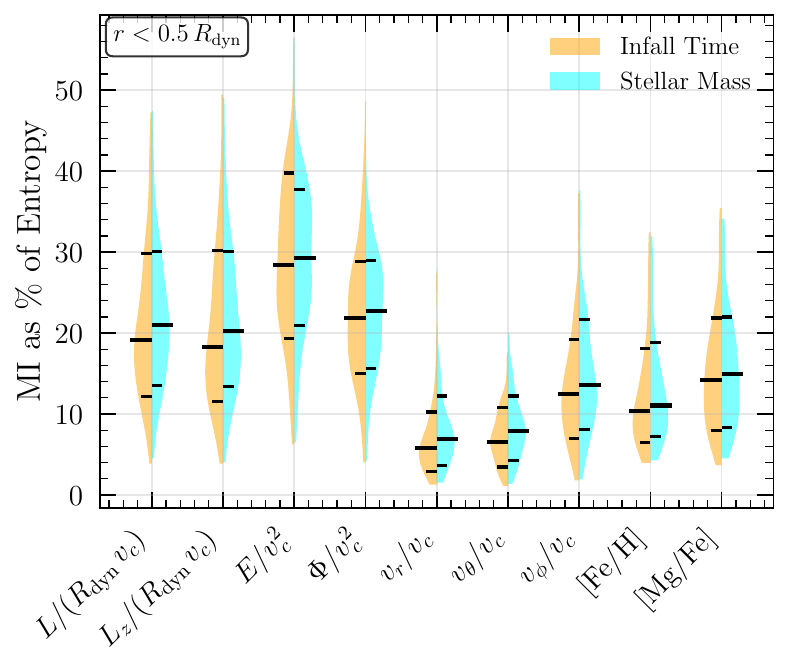}
    \caption{\gls{MI} between each feature and the infall time (orange) and stellar mass (cyan) of the merging satellite normalized by the entropy of the prediction variable. The  \gls{MI} is computed for each galaxy separately.
    The violin plots show the median, as well as the 16th and 84th percentiles of the normalized \gls{MI} per each variable as defined in Sec.~\ref{sec:features}.}
    \label{fig:feature_importance}
\end{figure}

To understand which features hold the most information about the merger history of the galaxy, we compute the \gls{MI} between each feature and the two main predictions (stellar mass and infall time), as shown in Fig.~\ref{fig:feature_importance}. In this figure, we limit the stellar sample to those within $r<0.5~R_{\rm{dyn}}$. Each galaxy's stellar distribution has a particular value for the \gls{MI}, and we show the 16th, 50th, and 84th percentiles across the 98 galaxies as black lines atop the overall distribution. The distributions of \gls{MI} values are broadly similar between the two prediction targets, but two competing effects make a precise comparison difficult. First, infall time is recorded at discrete snapshot intervals, which compresses the support of infall time and tends to inflate its normalized \gls{MI} relative to a continuous label. Second, stars from a single merger do not share a single infall time---different stars cross the virial radius at different snapshots depending on when they become unbound from the progenitor---whereas they do share a single progenitor stellar mass by construction. This within-merger spread in infall time acts as label noise that suppresses \gls{MI}, in the opposite direction of the discretization effect. Because the two biases push opposite ways and we have no clean way to subtract either, we report \gls{MI} for both predictions throughout and avoid drawing conclusions that depend on small differences between them. 

For the features considered in Fig.~\ref{fig:feature_importance}, the most informative across the entire sample are the (normalized) energy, potential, and angular momentum. In Fig.~\ref{fig:feature_importance}, we find that energy has a median \gls{MI} of $\osim 0.3$ across the sample, followed by the potential and angular momenta with medians of $\osim 0.2$. Energy and angular momentum are the most commonly used features to find substructure in the Milky Way~\citep[][]{1999MNRAS.307..495H,2000MNRAS.319..657H,BinneyTremaine:2008}, given that they are approximately conserved in a slowly evolving potential. Energy is conserved in the case of a static potential, while $L_z$ is conserved in the case of axisymmetry. Neither of these two quantities are expected to be exactly conserved through the galactic assembly process, however, information is retained in these features, suggesting their importance for understanding the history of the Milky Way.

Chemical abundances [Fe/H] and [Mg/Fe] have a median normalized \gls{MI} of $\lesssim 0.10$--0.15, with [Fe/H] being the less informative feature for both predictions. This is somewhat surprising given that chemical abundances are expected to be robust tracers of merger history, as they are intrinsic stellar properties unaffected by dynamical mixing. We caution, however, that this low \gls{MI} likely reflects limitations of the IllustrisTNG chemical enrichment model rather than a genuine feature of the Milky Way: the subgrid prescriptions in IllustrisTNG are known not to reflect the chemical abundance distribution of the Milky Way~\citep{2018MNRAS.473.4077P, 2018MNRAS.477.1206N}. Consistent with this interpretation, the same analysis applied to the FIRE-2 sample (Sec.~\ref{sec:fire_comparison}, and shown in Fig.~\ref{fig:feature_comparison}) yields systematically higher \gls{MI} in [Fe/H] and [Mg/Fe] by $\osim{}10$ percentage points, suggesting that the chemical channel carries more merger information than TNG50 alone implies. The values reported here for chemistry should therefore be treated as a lower bound on what is achievable in the real Galaxy.

Finally, the least informative features are the normalized velocities, with $v_r$ being the least informative component with a normalized \gls{MI} of $\osim 0.05$, followed by $v_\theta$ ($\osim 0.08$) and $v_\phi$ ($\osim 0.12$). As we will show later, this is because the velocity information is more easily washed out by phase mixing, and therefore it is not as robust as energy or angular momentum. %

\begin{figure*}[t]
    \centering
    \includegraphics[width=0.9\linewidth]{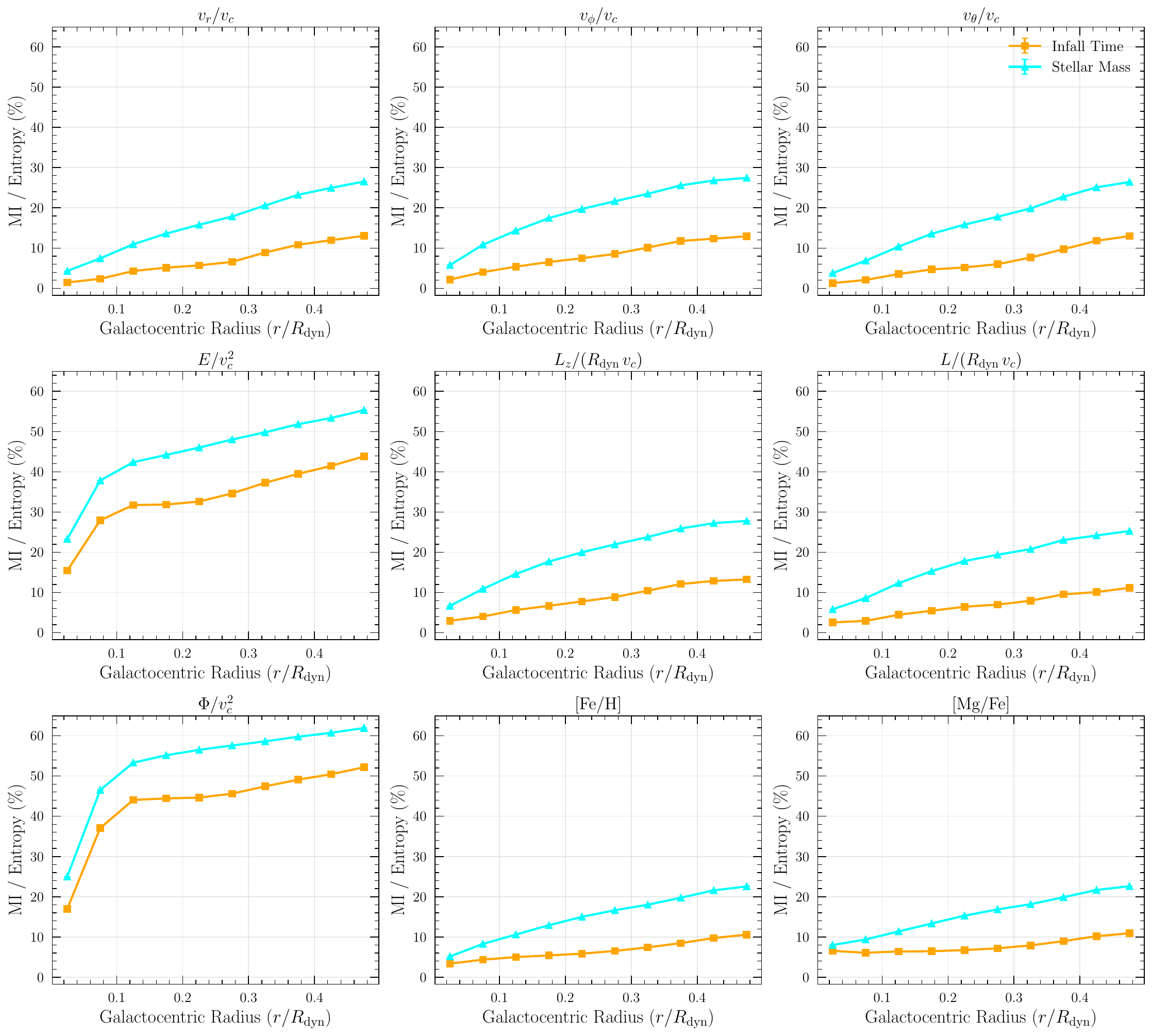}
    \caption{\gls{MI} of select features normalized by the entropy of the prediction variable in each bin, shown across multiple radial bins. Different features show distinct radial dependencies: energy and angular momentum information decrease toward the inner galaxy where dynamical times are shorter, while metallicity remains relatively constant across all radii. Note that these points include bootstrapped error bars, though they are too small to be visible given the large sample size of stars. \label{fig:radial_split}}
\end{figure*}

\subsection[Dependence of the MI on the Galaxy Properties]{Dependence of the \gls{MI} on the Galaxy Properties} \label{sec:galaxy_props}

In this section, we explore how the \gls{MI} between the features and the predictions depends on the properties of the stars, such as their radial position or the infall time of their progenitor. We find that the \gls{MI} is not uniform across the galaxy, but rather it depends on three key factors:
(1) \textit{Radial position}—stars in the inner galaxy experience shorter orbital times and more rapid phase mixing, leading to faster information loss (Sec.~\ref{sec:radial_dynamical}); (2) \textit{Infall time}—older mergers have had more time for their stars to complete many orbits and lose dynamical memory (Sec.~\ref{sec:temporal_evolution}); and (3) \textit{Merger mass}—major mergers induce violent relaxation that erases information on a single dynamical time, independent of subsequent mixing (Sec.~\ref{sec:merger_mass}).

\subsubsection{Radial Dependence and Dynamical Timescales} \label{sec:radial_dynamical}

The radial position of a star is correlated with its orbital timescale, which in turn controls the rate at which dynamical information about its merger origin is mixed away through gravitational interactions.

In Fig.~\ref{fig:radial_split}, we show the normalized \gls{MI} that select features hold regarding merger infall time (in orange) and merger stellar mass (in cyan) as a function of galactocentric radius, normalized by $R_\mathrm{dyn}$. To do so, we pool the stars from the 98 galaxies in the sample, and bin them by normalized radius. Then, for each bin, we compute the \gls{MI} and the entropy for the different variables considered. 
For each bin, we bootstrap the stars within each bin 100 times with replacement~\citep{Efron1979BootstrapMA}, and compute the \gls{MI} for each of the bootstrapped samples. The error bars shown in Fig.~\ref{fig:radial_split}, albeit small, are then computed as the 16th and 84th percentiles of the \gls{MI} across the 100 bootstrapped samples.\footnote{Given that \gls{MI} is positive definite, and that bootstrapping introduces some correlations between the values, we find that the 50th percentile value of the bootstrapping is always marginally higher than the original central value. To be on the conservative side, we report the 50th percentile value in these plots, although the trends are the same.} %

We observe a clear radial gradient in the information content of dynamical features. The potential $\Phi$, and to a slightly lesser extent the total $E$, encode the most information about the properties of the mergers. (Note that in a fixed radial bin, $\Phi$ is more informative than $E$ as it captures more of the small substructure differences, contrary to what was found when these quantities were considered globally in Fig.~\ref{fig:feature_importance}; we discuss the physical reason for this in Sec.~\ref{sec:temporal_evolution}). For both $E$ and $\Phi$, the amount of information encoded increases rapidly until $\osim 0.1~R_{\rm{dyn}}$, after which it continues to increase, albeit much more modestly. The remaining dynamical features (velocities and angular momentum) increase essentially linearly, with varying---but overall shallow---slopes. 

The information content of the chemical abundances [Fe/H] and [Mg/Fe] regarding the merger infall time remains relatively flat across all radial bins, consistent with their nature as intrinsic stellar properties that are immune to dynamical mixing. There is some information on the stellar mass encoded in the chemical abundances, which increases slightly with radius. This might be due to dynamical friction affecting different-mass galaxies differently, therefore inducing a dependence on the radial distribution~\citep{1943ApJ....97..255C}. It is important to also highlight that this is the information encoded in the \textit{ex-situ} star population, and therefore the chemical abundances might have a much smaller range of variation that if we were to include the \textit{in-situ} stars as well, which would likely increase the \gls{MI} of the chemical abundances, although not necessarily to disentangle the different mergers from each other.

This increasing trend of the dynamics and kinematics features reflects the fact that stars in the inner galaxy have completed many more orbits since their host satellite's infall, leading to greater phase mixing and loss of the initial dynamical signature. 
Additionally, within the dynamical friction framework, the most massive mergers sink to the inner galaxy on a dynamical time, and therefore lose their information on a dynamical time as well. The fact that the energy and potential show a stronger decrease in the inner galaxy than the other features is due to the fact that violent relaxation induced by the merger directly affects $\partial \Phi/\partial t$ (and subsequently $\partial E/\partial t$), which erases information on the much shorter dynamical time that is in the inner galaxy~\citep{1967MNRAS.136..101L}. We therefore clarify that there are two effects at play: the shorter dynamical times, and dynamical friction. We do not disentangle these effects in this section, but we do so in Sec.~\ref{sec:merger_mass} and Fig.~\ref{fig:feature_mass_comparison}.

\begin{figure}[t]
    \centering
    \includegraphics[width=0.9\linewidth]{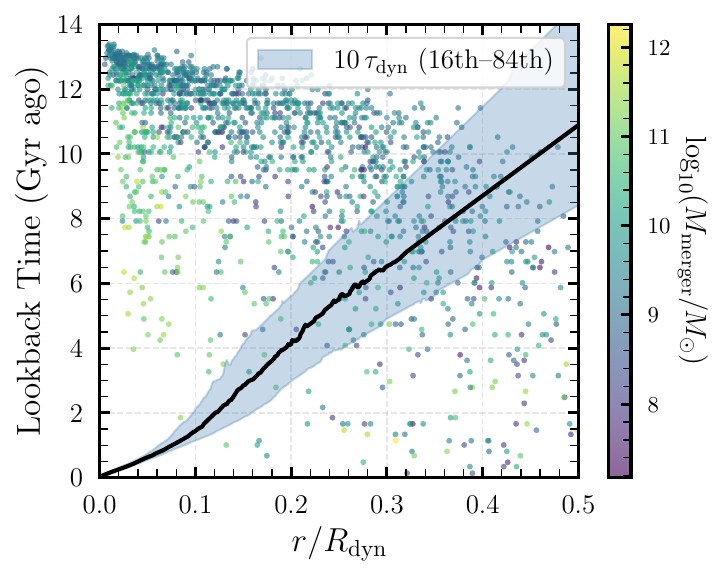}
    \caption{Infall time of all the mergers with at least 50 stars within $0.5~R_{\rm{dyn}}$ of the center of their host at present day, as a function of the average scaled radius of stars from the merger. 
    Overplotted is $10\times \tau_{\rm{dyn}}$, computed using the circular velocity as a function of scaled radius within each galaxy, with the mean across the 98 galaxies shown as a dark line, and the 16th--84th percentiles shown as a shaded region. 
    We caution that the computation of $\tau_{\rm{dyn}}$ is a simplification that assumes circular orbits, and therefore it is not accurate for highly radial orbits, but it serves as a useful proxy to understand the phase mixing timescales.
    The mergers above this line have had enough time to orbit their host many times and are considered virialized.
    The colorbar shows the mass of the different mergers, which shows 
    evidence for the effects of dynamical friction, as the most massive mergers tend to be further in for the same infall time. 
     \label{fig:dynamical_time}}
\end{figure}

\textbf{Dynamical Times as a Function of Radius:}
The radial trends observed in Fig.~\ref{fig:radial_split} can be understood to first order in terms of the dynamical timescale, which sets the rate of phase mixing. The dynamical time $\tau_{\rm{dyn}}$ at a given radius $r$ can be approximated as
\begin{equation}
\tau_{\rm{dyn}} = \frac{r}{v_{\rm{rot}}(r)},
\end{equation}
where $v_{\rm{rot}}(r)$ is the rotational velocity at that radius,
specifically defined as
\begin{equation} \label{eq:vc_r}
v_{\rm{rot}}(r) = \sqrt{v_\phi(r)^2 + v_\theta(r)^2}.
\end{equation}
This timescale represents the characteristic time it takes for a star to complete an orbit around the galaxy at radius $r$. Given that we only have access to the present-day properties of the stars, we compute $v_{\rm{rot}}(r)$ using the present-day velocities of the stars in each galaxy, and then compute $\tau_{\rm{dyn}}$ as a function of radius for each galaxy.

At small radii ($r \sim 10$~kpc), typical orbital times are a few hundred Myr, implying that stars have completed tens of orbits over a Hubble time. At large radii ($r \sim 50$--100~kpc), orbital times approach several Gyr, so stars may have completed only a handful of orbits since the formation of the Galaxy. The number of orbits completed since a merger's infall is roughly $N_{\rm{orb}} \sim t_{\rm{infall}} / \tau_{\rm{dyn}}$, and greater values of $N_{\rm{orb}}$ correspond to more thorough phase mixing.

To quantify when a merger should be significantly washed out, we adopt a conservative criterion: a merger is considered virialized if its infall time exceeds $10\times \tau_{\rm{dyn}}$ at the current radial position of its stars. It should be caveated that this is a simplifying assumption that orbits are close to circular. For extremely radial orbits, this estimate is not correct, but we will nevertheless use it as a proxy to understand phase mixing. In Fig.~\ref{fig:dynamical_time}, we show the infall time of all mergers with at least 50 stars within $0.5~R_{\rm{dyn}}$, plotted against the average scaled radius of stars in that merger. We overplot the $10\times \tau_{\rm{dyn}}$ threshold as a dark line, assuming the values from the simulations; specifically, for each \textit{scaled radial} bin, we compute the range of $10\times \tau_{\rm{dyn}}$ from the different simulations, and the individual $v_{\rm{rot}}(r)$ of the disk stars within each galaxy, interpolated as a function of $r$ (see Eq.~\ref{eq:vc_r} and Fig.~\ref{fig:vc}). %

\begin{figure*}
    \centering
    \includegraphics[width=0.95\linewidth]{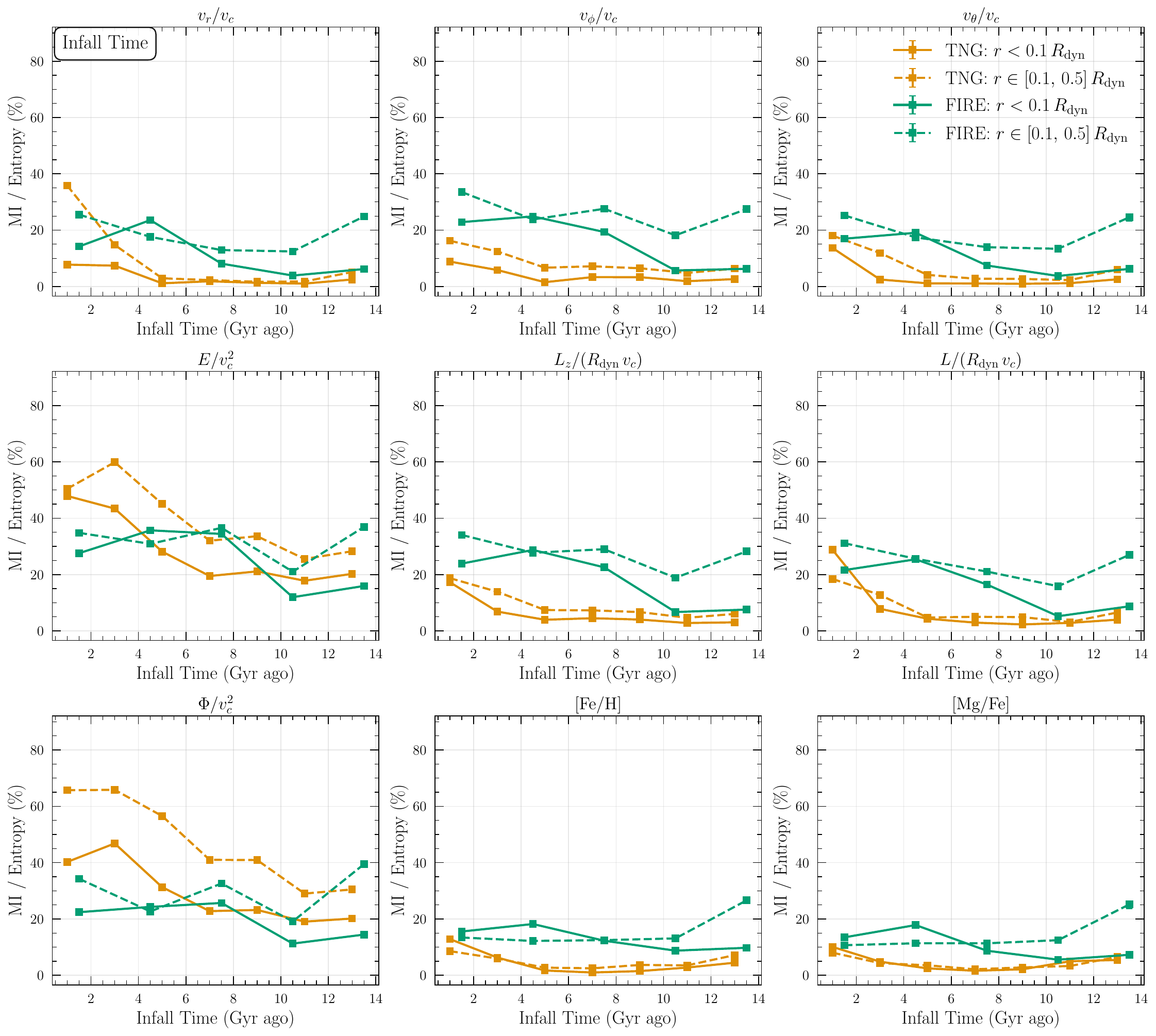}
    \caption{\gls{MI} of select features normalized by the entropy at each bin for the full TNG50 sample, as well as three FIRE-2 galaxies. We specifically focus on the inner galaxy ($r<0.1~R_{\rm{dyn}}$) and the outer galaxy ($r \in [0.1,0.5]~R_{\rm{dyn}}$) and the information as it pertains to computing the infall time (A parallel figure for the stellar mass is shown in Fig.~\ref{fig:feature_comparison_stellar_mass}). As expected from phase-mixing arguments, \gls{MI} decreases with increasing lookback time for all dynamical features, with the effect more pronounced in the inner galaxy where orbital times are shorter. Chemical abundances remain relatively constant across time bins, consistent with its role as an intrinsic tracer. Interestingly, the behavior does not vary much between the TNG50 and FIRE-2 samples, albeit with some normalization differences, suggesting that the information washout processes are robust across different galaxy formation models.
    \label{fig:feature_comparison}}
\end{figure*}

The majority of mergers lie well above this threshold, indicating that most mergers in our sample should be dynamically well-mixed by the present day. However, a non-negligible fraction of mergers—particularly those that fell in recently or whose stars now reside at large radii—lie below or near the threshold, suggesting that some dynamical memory of these events persists, and would be detectable in the present-day dynamics of the stars. 

\subsubsection{Temporal Evolution: Infall Times and Washout} \label{sec:temporal_evolution}

Information washout is fundamentally a time-dependent process: the
longer ago a merger occurred, the more orbits its debris stars have
completed, and the more thoroughly the initial phase-space signature
has been eroded by dynamical mixing and potential evolution.

We examine how \gls{MI} varies with the lookback time to
merger infall. Fig.~\ref{fig:feature_comparison} shows the normalized
\gls{MI} for select features as a function of lookback time,
separately for the inner galaxy (solid lines,
$r < 0.1\;R_{\mathrm{dyn}}$) and outer galaxy (dashed lines,
$r \in [0.1, 0.5]\;R_{\mathrm{dyn}}$). As points of comparison, we also show the lines for the mergers from the three FIRE-2 galaxies (see Sec.~\ref{sec:fire_comparison}), with the \gls{MI} for the inner galaxy shown with solid lines and the outer galaxy with dashed lines. The normalization follows the
procedure described in Sec.~\ref{sec:radial_dynamical}, dividing the
\gls{MI} by the marginal entropy of the target variable
(infall time or stellar mass) within each time bin.

\paragraph{Energy and gravitational potential}
The gravitational potential $\Phi$ and total specific energy
$E = \Phi + v^2/2$ retain the most merger information across all
lookback times. For recent mergers (within the past few Gyr), $\Phi$
preserves $\osim{}80\%$ of the available information in the outer
galaxy and $\osim{}50\%$ in the inner galaxy, with $E$ retaining
slightly less. That energy carries the most information is expected on
dynamical grounds: in a time-independent
potential, $E$ is an exact integral of motion, and even in the
time-varying, non-spherical potentials of our TNG50 sample, it remains
approximately conserved over many orbital periods. The \gls{MI} encoded in
$E$ therefore reflects the characteristic energy distribution imprinted
on debris at the time of stripping, which differs systematically with
satellite stellar mass and infall time---more massive satellites
deposit debris over a wider range of binding energies, and earlier
infall times produce debris on more tightly bound orbits due to
subsequent adiabatic contraction of the host.

The slight excess of \gls{MI} in $\Phi$ relative to $E$, most pronounced in the outer galaxy, reflects an asymmetry in how the two observables respond to a time-evolving host potential. Within a narrow radial bin, $E - \Phi = v^2/2$, so the comparison between $\mathcal{I}(E; Y)$, with $Y$ being the merger label, and $\mathcal{I}(\Phi; Y)$ amounts to asking whether the kinetic-energy contribution to $E$ adds discriminative information or noise. In a static potential, $E$ is an exact integral of motion and would be the optimal merger tracer. In cosmologically evolving potentials, however, individual stellar energies drift secularly by $\Delta E = \int \partial_t \Phi(\vec{r}(t), t)\, dt$ along each orbit \citep{Gomez2010}, with a magnitude set by the star's specific orbital and accretion history. This accumulated, realization-dependent drift adds noise to $E$ that has no analog for $\Phi$, since $\Phi(\vec{r}, t)$ at a star's present position is determined entirely by the current mass distribution (and thus any present substructure) and is not a time-integrated quantity. The drift is most consequential in the outer galaxy, as can be seen in Fig.~\ref{fig:feature_comparison}, where orbits sample radii at which the potential continues to be reshaped by accretion and recent mergers, while in the inner galaxy the potential, being far deeper, is sufficiently stable that $E$ remains nearly as informative as $\Phi$. The smallness of the gap throughout confirms that secular potential evolution is mild relative to dynamical timescales over most of our sample.

The systematic decrease of \gls{MI} in both $E$ and $\Phi$ with increasing
lookback time reflects the cumulative effect of potential evolution:
older mergers have experienced a longer history of mass accretion, structural
rearrangement, and perturbation from subsequent infalling satellites,
all of which gradually shuffle stellar energies away from their values
at the time of stripping. This is genuine information loss---the
mapping from initial merger parameters $(M_\star,\; t_{\rm infall})$ to
present-day $E$ becomes progressively more stochastic across the
ensemble of host galaxies---and operates at any measurement
resolution, distinguishing it from the phase-mixing effects discussed
below.

\paragraph{Angular momenta}
The angular momenta $L$ and $L_z$ carry substantially less information
than $E$ or $\Phi$ at all lookback times, retaining $\osim{}20\%$
of the available information for recent mergers in the outer galaxy
and significantly less in the inner galaxy. Notably, $L$ and $L_z$
show nearly identical \gls{MI} values across all lookback times and radial
bins in TNG50, differing by only a few percent.

The low information content has its origin in the merger process
itself. Satellite infall is fundamentally radial, and as shown by
\citet{2022ApJ...926..203V}, the orbits of massive satellites undergo
further radialization during the merger: self-gravity torques from
both the host's dipolar response and the satellite's own tidally
stripped debris preferentially reduce the orbital angular momentum,
particularly for mass ratios $\gtrsim 1{:}10$ and 
eccentric orbits. The debris is thus deposited on
highly radial orbits whose angular momentum distribution is compressed
toward low values, reducing the dynamic range available to encode
differences between merger parameters. This places a ceiling on the
information that any angular momentum component can carry,
independently of how well it is subsequently conserved.

The near-identical \gls{MI} of $L$ and $L_z$ is itself informative about
the geometry of the host potential. $L_z$ is conserved under
axisymmetry alone, while conservation of $L$ additionally requires
spherical symmetry. The fact that both quantities retain the same
information implies that the potentials of our TNG50 Milky Way
analogues are close to spherical over the radial range and timescales
relevant to merger debris. %

Interestingly, however, in Fig.~\ref{fig:feature_comparison_stellar_mass}, in which we study the \gls{MI} of the different features to the stellar mass of the progenitor, we find that $L_z$ retains much more information than $L$ in the FIRE-2 sample, with a strong decay over the last $\osim 8$~Gyr. This difference emphasizes the axisymmetric nature of the FIRE-2 potentials, where disks are more pronounced than in TNG50. Additionally, FIRE-2 is run at slightly higher resolution and is therefore able to resolve finer structures, including that of the disk.

\paragraph{Radial velocity}
The individual velocity components, unlike the dynamical features, retain information for significantly less time than the orbital integrals, which is expected: unlike energy or angular momentum, the velocities $(v_r, v_\theta, v_\phi)$ are not approximate constants of motion and are therefore directly modified by the time-varying gravitational field during and after the merger. Two distinct regimes emerge from Fig.~\ref{fig:feature_comparison}. First, $v_r$ exhibits a sharp decline in \gls{MI}, most pronounced in the outer galaxy, where the information is consistent with the noise floor beyond $\osim 5$~Gyr ago. Second, the tangential components $v_\theta$ and $v_\phi$ behave qualitatively differently: $v_\theta$ decays moderately---faster in the inner galaxy where meridional oscillation frequencies are higher---while $v_\phi$ remains nearly constant across all time bins for TNG50, while falling for FIRE-2, but mirroring the behavior of $L_z$ in both cases. We discuss the physical origin of each regime below.

To interpret the sharp decrease in the information encoded in $v_r$,
we first stress an important distinction: \textit{a large change in the value of
$v_r$ over time does not, by itself, imply a loss of information.} If the
radial force $F_r = dv_r/dt$ modified every star's radial velocity in
a way that depended deterministically on the merger parameters
$(M_\star,\; t_{\rm{infall}})$, the mapping from merger properties to
present-day $v_r$ would remain one-to-one, and the \gls{MI}
would be preserved. Information is destroyed only when the mapping
becomes many-to-one---i.e.\ when mergers with the same
$(M_\star,\; t_{\rm{infall}})$ produce systematically different
present-day $v_r$ distributions. Two mechanisms drive this
stochasticity.

First, \textbf{orbit-dependent center-of-mass shifts:} as
\cite{2022ApJ...926..203V} show, during a merger the barycenter of
the host halo is displaced from its equilibrium position, inducing a
time-dependent, realization-specific radial force on all halo stars.
The magnitude and direction of this displacement depend on the full
three-dimensional orbital geometry of the infalling satellite---its
eccentricity, inclination, and pericentric distance---none of which
are captured by $(M_\star,\; t_{\rm{infall}})$ alone. Two mergers with
the same mass and infall time but on, e.g., a near-circular versus a
highly radial orbit produce qualitatively different barycenter
trajectories and therefore different ranges of $v_r$ for the
stellar debris. Across the ensemble of host galaxies, this orbital
diversity renders the mapping from merger labels to present-day $v_r$
many-to-one, erasing \gls{MI}.

Second, \textbf{causal dependence on prior merger history:} even for
mergers on similar orbits, the host halo's mass distribution at the
time of infall depends on the sequence and geometry of all previous
accretion events. Two hosts that experienced different earlier mergers would
present different potentials to the same $(M_\star,\; t_{\rm{infall}})$
event, leading to different barycenter displacements and therefore
different $v_r$ evolution. This source of stochasticity cannot be
removed by conditioning on the satellite's orbital parameters; it is
an irreducible consequence of the diversity of assembly histories
across the galaxy population, and it compounds with the
orbit-dependent mechanism above. Together, these two effects explain
why $v_r$ loses information more rapidly than the orbital integrals:
unlike $E$ or $L_z$, whose evolution is governed by comparatively
smooth, potential-averaged quantities, $v_r$ is sensitive to the
instantaneous, realization-specific gravitational field, which varies
stochastically across the ensemble.

\paragraph{Angular velocities}
The tangential components $v_\theta$ and $v_\phi$ retain
substantially less \gls{MI} overall ($\osim 20\%$ for recent
mergers) than the radial velocity or the orbital integrals, but their
temporal behavior differs markedly from $v_r$: neither component shows the
sharp, complete washout seen in $v_r$, and $v_\phi$ in particular
remains nearly constant with lookback time. The low overall
information content can be understood from the geometry of the merger
interaction. During infall, the dominant gravitational force between
the satellite and the host is radial; the tangential velocity of the
debris is therefore set primarily by the satellite's pre-infall
orbital angular momentum, diluted by the satellite's internal velocity
dispersion $\sigma_{\rm sat}$. For a satellite on an orbit with
tangential velocity $v_{\rm orb,\perp}$, the merger-specific signal
in $v_\theta$ or $v_\phi$ for individual debris stars scales as
$v_{\rm orb,\perp} / \sqrt{v_{\rm orb,\perp}^2 + \sigma_{\rm
sat}^2}$: massive satellites with large $\sigma_{\rm sat}$ relative
to their orbital velocity contribute debris whose tangential
velocities are dominated by the satellite's internal kinematics rather
than the orbit, reducing the correlation with merger properties. This
effective signal-to-noise ratio is generically lower for the
tangential components than for $v_r$, where the radial force during
infall imprints a coherent, orbit-dependent signature on all debris
stars. Below we discuss $v_\phi$ and $v_\theta$ separately, as they
are governed by distinct physics.

We first consider $v_\phi$, which at fixed galactocentric radius is a
direct proxy for $L_z = R\,v_\phi$. As shown in
Fig.~\ref{fig:feature_comparison}, $v_\phi$ and $L_z$ carry nearly
identical \gls{MI} at every time bin, for both TNG50 and FIRE-2, confirming that the
information content of $v_\phi$ is governed by the physics of $L_z$.
In an axisymmetric potential, $L_z$ is exactly conserved; information
about the merger that set its value is therefore preserved
indefinitely. Real galactic potentials are not perfectly axisymmetric,
and mergers themselves break the symmetry: the self-gravity of the
infalling satellite and the reflex motion of the host halo exert
non-axisymmetric torques that change $L_z$ over
time, especially during pericentric passages~\citep{2022ApJ...926..203V}. 
The change
$\Delta L_z$ correlates strongly with the merger's mass,
orbit, and infall time, preserving to some degree the one-to-one character of the
mapping from merger parameters to present-day $L_z$. By contrast, the
barycenter displacements that dominate changes in $v_r$ depend on the
full three-dimensional geometry of the interaction, which varies
stochastically across realizations (see above). Once the satellite is
fully disrupted and the host potential relaxes back toward
axisymmetry, $L_z$ is again approximately conserved, and no further
information loss occurs. The net result is that $v_\phi$ and $L_z$
show only mild variation in \gls{MI} with lookback time,
declining primarily during the brief, non-axisymmetric phase of the
merger itself rather than accumulating losses over many orbital
periods.

In contrast, the \gls{MI} between $v_\phi$ and stellar mass, as shown in~Fig.~\ref{fig:feature_comparison_stellar_mass}, shows a steady decline with lookback time, which is mirrored by the \gls{MI} of $L_z$. This suggests that the torques that change $L_z$ during the merger are more effective at erasing information about stellar mass than infall time in FIRE-2. This could be due to the fact that more massive mergers experience stronger torques and therefore have their $L_z$ changed more significantly, while the infall time is not as strongly correlated with the strength of the torques. Additionally, we are only limited to three galaxies in the FIRE-2 sample, which might not be enough to capture the full diversity of merger histories and therefore the full range of torques experienced by the different mergers. 

Second, we discuss the information content of $v_\theta$. Unlike
$v_\phi$, which at fixed position is a direct proxy for the conserved
quantity $L_z$, the polar velocity $v_\theta$ depends on where a star
sits in its meridional oscillation cycle---i.e.\ on the conjugate
angle variables $(\theta_R,\theta_z)$ as well as on the actions~\citep{BinneyTremaine:2008}.
Three processes contribute to the decay of information in $v_\theta$.

First, \textbf{meridional phase mixing:} debris stars stripped from the
same satellite have a spread in orbital frequencies set by the
satellite's internal velocity dispersion. Over time, their meridional
phases decohere, so that stars initially at similar points in their
$(R,z)$-plane oscillation become uniformly distributed in orbital
phase. At a fixed spatial location, $v_\theta$ then reflects only the
distribution of actions, not the initial phase coherence that carried
merger-specific information. The decoherence timescale scales with the
local dynamical time, so it proceeds faster in the inner galaxy.
In a perfectly integrable, static potential, this phase mixing is
strictly a finite-resolution effect: the information migrates to
progressively finer phase-space structure rather than being destroyed,
and a sufficiently fine-grained estimator would recover it. Our \gls{MI}
estimator, operating at finite sample size and bin resolution,
registers this as information loss---which is also the
observationally relevant limit.

Second, \textbf{secular time evolution of the potential:} the host galaxy's
potential is not static. During the active merger phase, violent
fluctuations in the gravitational potential alter stellar orbits
stochastically, modifying the actions themselves (including $E$ and $L_z$) in a
realization-dependent manner
\citep[][]{1967MNRAS.136..101L,2022ApJ...926..203V}. Even after the merger
concludes, continued mass accretion and the growth of baryonic
structures such as the disk and bar slowly evolve the potential, causing secular
diffusion, or a slow stochastic evolution, of the stellar actions~\citep{1988MNRAS.230..597B}. Unlike phase mixing, this process destroys
information at any resolution, because the mapping from initial merger
parameters to present-day actions becomes many-to-one across the
ensemble of host galaxies.

Third, \textbf{chaotic mixing:} realistic galactic potentials are not always integrable, and thus not
all orbits admit well-defined integrals of motion. Stars on
chaotic orbits diffuse through the accessible meridional phase space
at a rate that depends sensitively on initial conditions, scrambling
$v_\theta$ faster than regular phase mixing. The fraction of chaotic
orbits is expected to be higher in the inner galaxy, where the
potential departs more strongly from a simple separable form,
contributing to the steeper radial dependence of $v_\theta$
information decay that we observe.

\paragraph{Chemical abundances}
[Fe/H] and [Mg/Fe] show little dependence on lookback time in either radial bin, remaining roughly constant across all time intervals, although with an interesting increase for the oldest stars. Chemical abundances are set at the time of star formation and are not affected by dynamical processes, so they retain their information about merger properties regardless of how long ago the merger occurred. The increase in \gls{MI} for the oldest stars could be due to the fact that these stars formed in a more chemically primitive environment, where differences in metallicity between mergers were more pronounced, making it easier to distinguish them based on their chemical signatures.
Of course, one needs to be wary of the numerical assumptions built into IllustrisTNG~\citep{2018MNRAS.473.4077P}. To that end, we compare the results of the chemical abundances in TNG50 to those of the three FIRE-2 galaxies, which are simulated at higher resolution and with different subgrid physics. We find that the chemical abundances in FIRE-2 show a similar behavior to those in TNG50, albeit retaining slightly higher information. This suggests that the information content of chemical abundances is robust across different galaxy formation models, although the specific values may differ due to differences in the implementation of star formation and feedback processes.

\subsubsection{Combined Effects: Infall Time and Current Radius} \label{sec:infall_radius_2d}

\begin{figure*}
    \centering
    \includegraphics[width=0.95\linewidth]{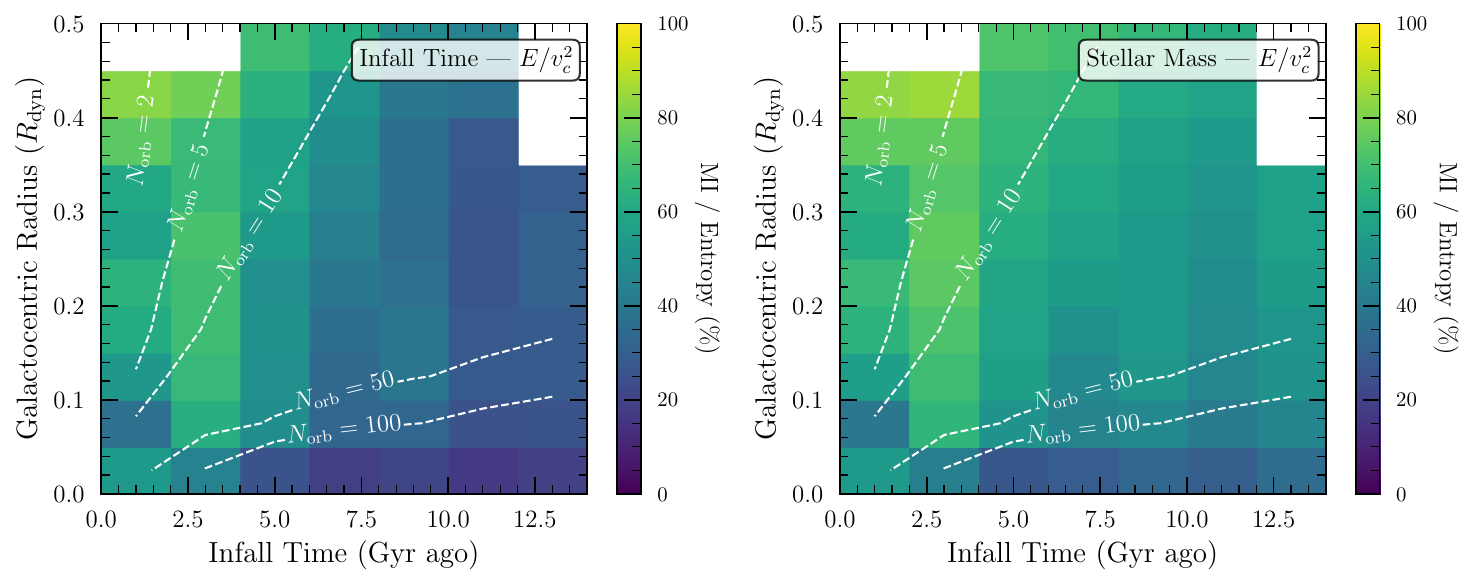}
    \caption{The normalized \gls{MI} per bin for the energy and inferring the infall time (Left) and stellar mass (Right), shown as a function of the infall time and current normalized galactocentric radius. High \gls{MI} (yellow) indicates strong information retention; low \gls{MI} (dark blue) indicates washout. Overlayed is the number of orbits $N_\mathrm{orb}$ per bin, assuming Eq.~\ref{eq:n_orb} per time and spatial bins. White bins include either less than 50 stars or fewer than two mergers. The highest information content is found for recent mergers (small lookback times) at large radii, where stars have completed few orbits. The lowest information is for old mergers at small radii, where extensive phase mixing has occurred.  
    \label{fig:infall_radius_2d}}
\end{figure*}

We have so far discussed the dependence of the dissipation of \gls{MI} as a function of time since infall and position. These two factors are not independent; they combine to determine the total number of orbits completed since infall. A star that fell in at lookback time $t_{\rm{infall}}$ and currently resides at radius $r$ has completed approximately
\begin{equation} \label{eq:n_orb}
N_{\rm{orb}} \approx \frac{t_{\rm{infall}}}{\tau_{\rm{dyn}}(r)} \approx \frac{t_{\rm{infall}} \, v_c(r)}{r}
\end{equation}
orbits since its host satellite merged. The degree of phase mixing—and hence the loss of dynamical information—scales primarily with $N_{\rm{orb}}$. Of course, this is the idealized picture; in reality, the orbital history can be more complex, with stars migrating radially over time, and mergers inducing non-spherical potentials that can enhance mixing. However, $N_{\rm{orb}}$ provides a useful first-order framework for understanding information washout.

To visualize this combined dependence, we show in Fig.~\ref{fig:infall_radius_2d} two-dimensional maps of the normalized \gls{MI} of the energy $E$, as a function of both infall time ($x$-axis) and current galactocentric radius ($y$-axis). The left panel shows the normalized \gls{MI} between the energy and the infall time, while the right panel shows the \gls{MI} between the energy and the stellar mass. Overlayed over both panels are contours of constant $N_{\rm{orb}}$ computed using Eq.~\ref{eq:n_orb} for each bin in infall time and radius, adopting the mean circular velocity profile across simulations at each specific radius. 

The 2D structure reveals a diagonal trend: the highest \gls{MI} (yellow regions) occurs in the upper-left corner, corresponding to recent mergers ($t_{\rm{infall}} \lesssim 2$~Gyr ago) at large radii ($r \gtrsim 0.35~R_{\rm{dyn}}$). In this regime, stars have completed only a few orbits ($N_{\rm{orb}} \sim 1$--2), and their phase-space distribution retains a strong memory of the merger event. Conversely, the lowest \gls{MI} (dark blue regions) occurs in the lower-right corner, corresponding to old mergers ($t_{\rm{infall}} \gtrsim 8$~Gyr ago) at small radii ($r \lesssim 0.1~R_{\rm{dyn}}$). Here, stars have completed $\mathcal{O}(100)$ orbits, and phase mixing has thoroughly erased the initial dynamical signature.

The transition between high and low \gls{MI} follows contours of roughly constant $N_{\rm{orb}}$, at least at first order. %
This suggests that the fundamental parameter controlling washout is not infall time or radius alone, but the number of orbits—or equivalently, the ratio $t_{\rm{infall}} / \tau_{\rm{dyn}}$. 

However, there are some deviations from this simple scaling. For instance, the \gls{MI} for the most recent mergers show a more complex structure, with non-monotonic changes to the \gls{MI} in this infall time bin  ($t_{\rm{infall}} \sim 0$--2~Gyr). We attribute this to the fact that these mergers are more massive and sink faster to the center of the potential well due to dynamical friction~\citep{1943ApJ....97..255C}, as can be seen in Fig.~\ref{fig:dynamical_time}, where the most massive mergers are found within 0.1~$R_{\rm{dyn}}$. This will be discussed in more detail in Sec.~\ref{sec:merger_mass}. 

The 2D maps in Fig.~\ref{fig:infall_radius_2d} also reveal that old mergers can retain some information if their stars currently reside at very large radii. For instance, a merger from 10~Gyr ago still shows moderate \gls{MI} ($\osim 0.5$--0.6) for stellar mass for stars at $r \sim 0.3~R_{\rm{dyn}}$. This suggests that the outermost halo is a particularly valuable region for reconstructing the early merger history of the Galaxy, as it preserves dynamical signatures that have been erased elsewhere.

Interestingly, the \gls{MI} for inferring stellar mass shows a somewhat different pattern than that for infall time. While both exhibit the same general diagonal trend, stellar mass retains more information at intermediate radii and lookback times. This can be physical or numerical: it may reflect the fact that stellar mass is a more global property of the merger event (indicative of its total mass and intrinsic velocity dispersion), whereas infall time is more sensitive to the detailed orbital history and phase-space structure of the debris, which is more easily erased by mixing. However, we warn that this could also be a numerical artifact arising from the finite snapshot frequency, which would more strongly affect the noisier infall time estimate.

\subsubsection{Merger Mass Effects} \label{sec:merger_mass}

\begin{figure*}
    \centering
    \includegraphics[width=0.95\linewidth]{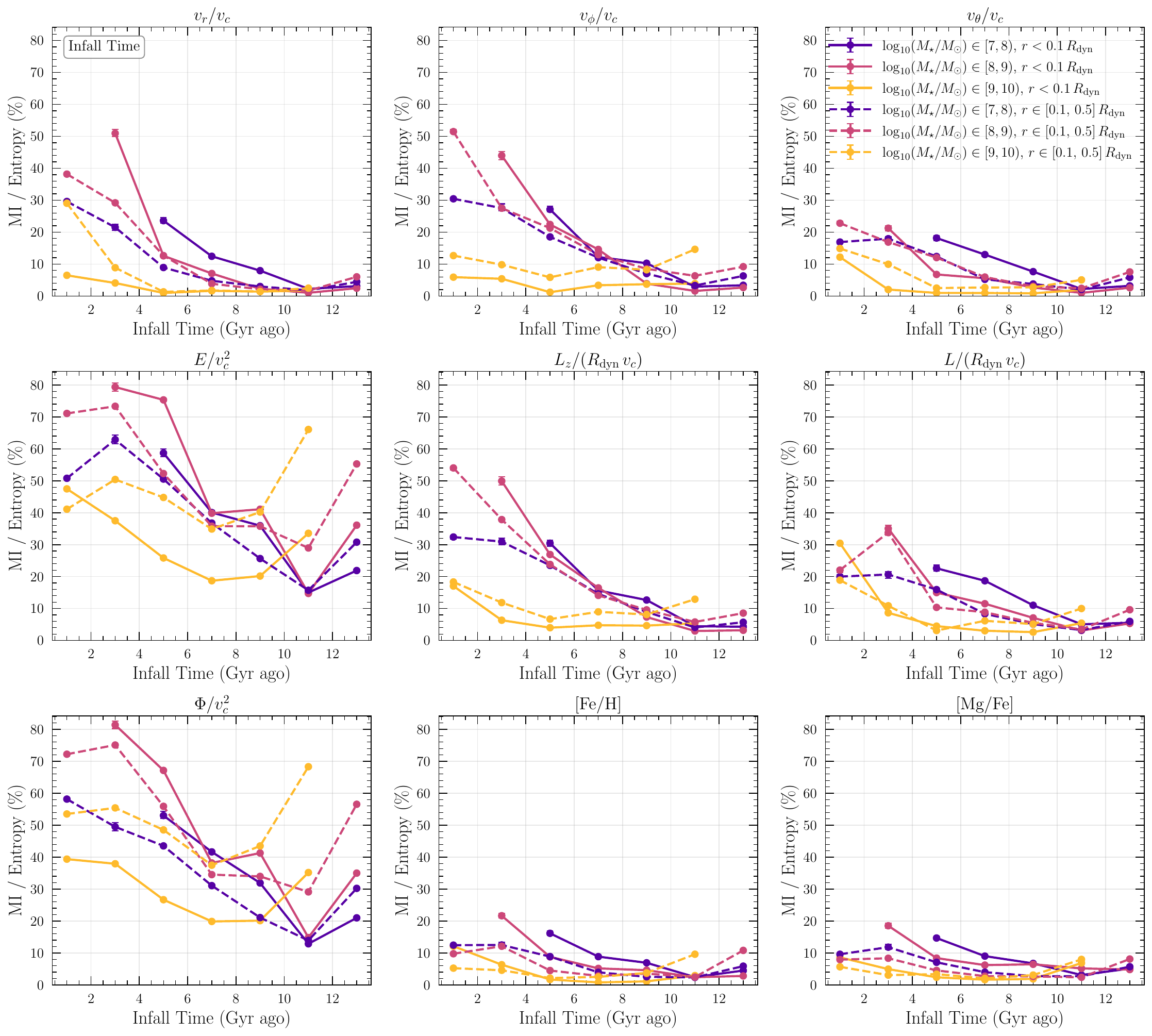}
    \caption{\gls{MI} of select features normalized by the entropy at each bin as compared to the infall time. Different colors correspond to different bins in the stellar mass of the merger, as indicated in the legend. The \gls{MI} is shown for the inner galaxy (solid lines, $r<0.1~R_{\rm{dyn}}$) and outer galaxy (dashed lines, $r \in [0.1,0.5]~R_{\rm{dyn}}$). Similar figure but for the stellar mass can be found in the Appendix as Fig.~\ref{fig:feature_mass_comparison_stellar_mass}. \label{fig:feature_mass_comparison}}
\end{figure*}

Up to this point, we have treated infall time and present-day radius
as the primary axes controlling information retention. The mass of
the merger is a third, physically distinct axis. Naively, more massive
satellites contribute more debris stars, which should improve the
statistical discriminability of their phase-space signature. Working
against this, however, massive satellites drive stronger perturbations
in the host potential during infall, and the resulting violent
relaxation scrambles stellar energies on a few dynamical timescales
\citep{1967MNRAS.136..101L}. These two effects compete, and their relative
importance determines whether mass aids or hinders merger
reconstruction.

Figure~\ref{fig:feature_mass_comparison} shows normalized mutual
information as a function of infall time, split into three stellar
mass bins at infall---low ($10^7 \leq M_\star/\Msun < 10^8$),
intermediate ($10^8 \leq M_\star/\Msun < 10^9$), and high
($10^9 \leq M_\star/\Msun < 10^{10}$)---and further split by
galactocentric radius (solid: $r < 0.1\,R_{\rm dyn}$; dashed:
$r \in [0.1, 0.5]\,R_{\rm dyn}$).

Across nearly all features, the high-mass bin retains the
\emph{least} information, not the most, indicating that violent
relaxation outweighs the larger stellar sample size for massive mergers.
The effect is most pronounced for the kinematic components
$v_r$, $v_\phi$, $v_\theta$, and the angular momenta $L_z$ and $L$: for
the most massive bin, these features carry essentially zero
information beyond $\osim 5$~Gyr ago, while lower-mass
mergers retain $\osim 15$--$25\%$ of the available information at
comparable lookback times. The chemistry panels show the same
ordering at all times: the low-mass bin holds a stable
$\osim 5$--$15\%$ \gls{MI} floor in both [Fe/H] and [Mg/Fe], while the
high-mass bin sits near $5\%$ throughout. This chemical
distinguishability of low-mass mergers reflects their characteristically
low metallicities and elevated $\alpha$-abundances---consequences of
early, short star-formation histories~\citep{Tolstoy2009}---which place them in regions of chemical space that are more sparsely populated than those occupied by higher-mass galaxies with more extensive merger and enrichment histories.

The energy and potential panels depart from this clean ordering in
two ways that are worth noting. First, at recent infall times
($\lesssim 4\;\mathrm{Gyr}$), the intermediate-mass bin actually
carries the highest \gls{MI} in $E$ and $\Phi$. This suggests that for
recent mergers, the larger stellar populations of intermediate-mass satellites do improve statistical resolution of the energy
distribution---violent relaxation has not yet had time to erase the
signal. Second, the high-mass bin in $E$ and $\Phi$ shows a
conspicuous rise at the oldest infall times ($\gtrsim 11\;\mathrm{Gyr}$),
reaching \gls{MI} comparable to the recent-infall values. We interpret this
as the signature of ancient massive mergers contributing
substantially to the construction of the host potential itself: their
debris no longer constitutes a distinguishable substructure, but the
overall energy distribution of stars in the host still reflects the
mass and epoch of these foundational events. A similar recovery is
not seen in the kinematic or angular momentum features, consistent
with the idea that what survives in this regime is the gross shape
of the potential rather than any coherent orbital structure.

Turning to the spatial dependence, the contrast between inner and
outer galaxy is generally smaller than the contrast between mass
bins, but it carries an interesting wrinkle. For the kinematic components $v_r$, $v_\phi$, $v_\theta$ in the low-mass
bin, the \emph{inner} galaxy is slightly more informative than the
outer. We speculate that this reflects a selection effect and a smaller sample of galaxies: low-mass
debris that has reached the inner galaxy at late times has done so
on relatively radial orbits, while outer-galaxy low-mass debris is dynamically
cooler and less distinguishable from the ambient halo population. A
more detailed investigation of this trend is deferred to future work.

\subsubsection{Effect of the Orbital Parameters of the Mergers} \label{sec:orbital_params}

\begin{figure*}[t]
    \centering
    \includegraphics[width=1\linewidth]{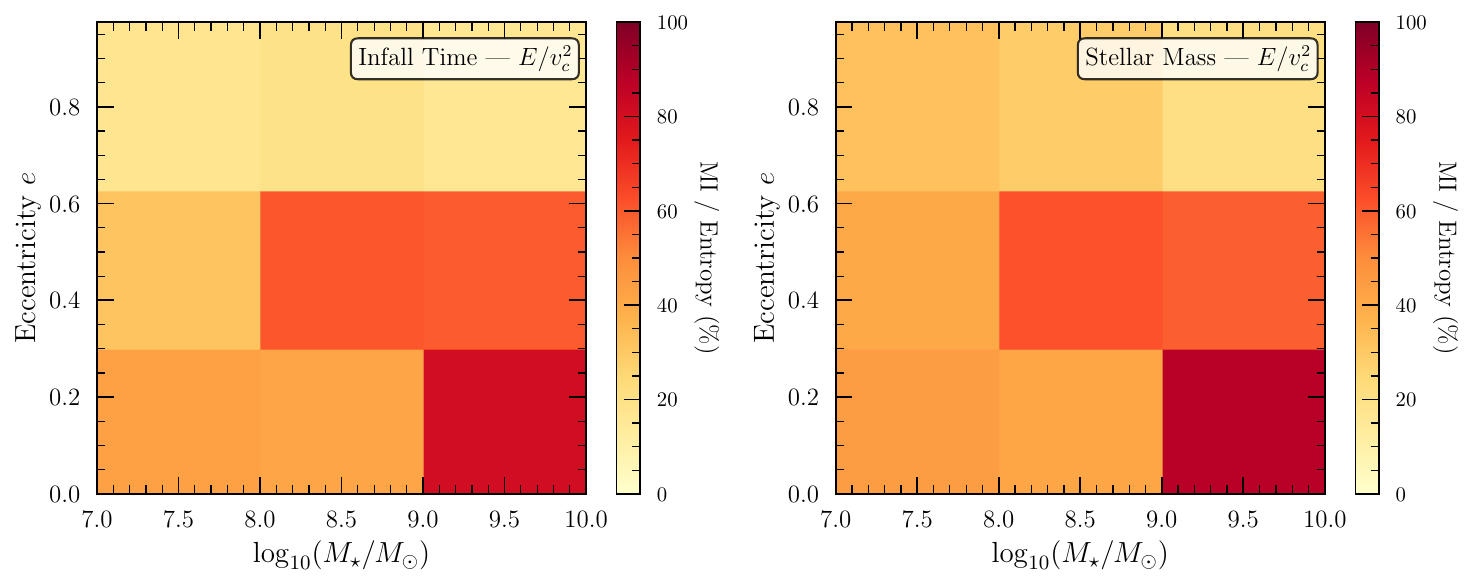}
    \caption{\gls{MI} of the energy and inferring the infall time (Left) and stellar mass (Right), shown as a function of the eccentricity and stellar mass of the merger. High \gls{MI} (red) indicates strong information retention; low \gls{MI} (yellow) indicates washout. The highest information content is found for mergers with low eccentricity that contribute the highest number of stars, while the lowest information is for mergers with high eccentricity, and is independent of the stellar mass. 
    \label{fig:orbital_params_mass_split}}
\end{figure*}

Another axis that plausibly affects information retention is the
orbital geometry of the infalling satellite. \citet{2022ApJ...926..203V}
showed that satellites on more radial orbits experience stronger
pericentric shocks and more efficient angular momentum loss through
self-gravity torques and debris self-friction, with the radialization
mechanism being most effective for initially eccentric, massive
mergers. Each pericentric passage drives violent fluctuations in the
host potential that scramble stellar energies through incomplete
violent relaxation~\citep{BinneyTremaine:2008,2013MNRAS.430..121P}, so we expect mergers on more eccentric orbits to
deposit debris that loses information faster than debris from
gentler, more tangential encounters of comparable mass.

To test this expectation, we estimate the orbital eccentricity of
each merger from its pre-infall trajectory in the simulation. For
each satellite identified by \textsc{SubFind}, we track its galactocentric
distance from the host across all snapshots in which it remains a
bound subhalo, and identify the apocenter and pericenter of its final
orbit before disruption as the maximum and minimum distances recorded
within this window. The eccentricity is then computed as $e = (r_{\rm apo} - r_{\rm peri}) / (r_{\rm apo} + r_{\rm
peri})$. We warn that this estimate is limited by the $\osim 150$~Myr snapshot
cadence of TNG50 \citep{2019ComAC...6....2N}: %
the
true pericenter and apocenter may fall between snapshots, biasing
$r_{\rm peri}$ high and $r_{\rm apo}$ low and therefore underestimating
the true eccentricity. These limitations preclude the determination of precise merger
eccentricities, but they preserve the relative ordering of mergers
from circular to radial to first order, which is sufficient for the
population-level trends we examine.

Figure~\ref{fig:orbital_params_mass_split} shows the \gls{MI} of
$E/v_c^2$ with infall time (left) and stellar mass (right) as a
function of merger eccentricity and stellar mass. We restrict the
analysis to energy alone because this observable---as well as 
$\Phi$---retains information when averaging over all radii and 
infall times, as established in Fig.~\ref{fig:feature_comparison}.
The other dynamical features carry too little information
across the full range of merger parameters to resolve eccentricity-dependent trends.

The figure reveals a clear three-regime structure. At high
eccentricity ($e \gtrsim 0.6$), the \gls{MI} is uniformly low
($\lesssim 30\%$) across all stellar mass bins---eccentricity dominates
and erases any mass-dependent signal. At intermediate eccentricities
($0.3 \lesssim e \lesssim 0.6$), the \gls{MI} is moderate and shows only a
mild mass dependence. At low eccentricities ($e \lesssim 0.3$), a
strong mass dependence emerges: the most massive mergers
($\log M_\star/\Msun \sim 9.5$) reach $\osim 85\%$ \gls{MI}, while the
lowest-mass bin retains only $\osim 50\%$.

This pattern indicates that mass and eccentricity do not act as
independent axes, but rather interact: eccentricity sets a ceiling on
how much information can survive the merger, while mass governs how
much of the surviving signal is statistically recoverable. Physically,
the eccentricity ceiling reflects the violent relaxation mechanism
discussed in Sec.~\ref{sec:temporal_evolution}: highly radial orbits
drive strong pericentric shocks and rapid potential fluctuations that
scramble stellar energies on a few dynamical timescales, washing out
the merger signature regardless of how many stars the satellite
contributed~\citep{JeanBaptiste2017}. Once eccentricity is low enough that violent relaxation
is mild, two complementary effects allow more massive mergers to
retain proportionally more information: (i) larger satellites
contribute more debris stars, providing finer statistical resolution
of the underlying energy distribution; and (ii) more massive
satellites have larger intrinsic velocity dispersions, so
their debris occupies a wider, more distinguishable range of energies
even before any post-merger evolution. Our current data cannot
distinguish these two contributions, but both are consistent with the
observed pattern.

\begin{figure*}
    \centering
    \includegraphics[width=0.95\linewidth]{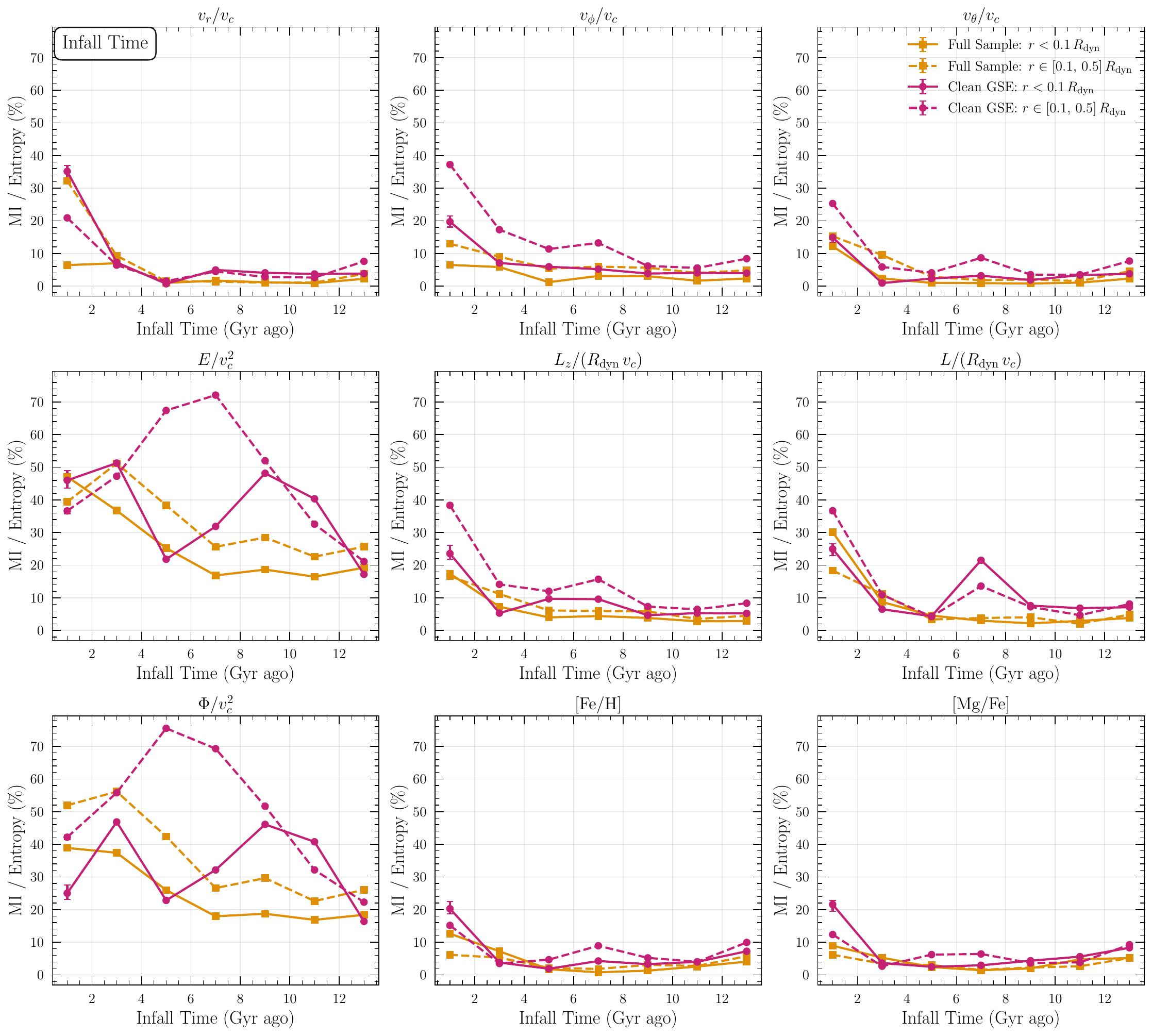}
    \caption{\gls{MI} of select features normalized by the entropy at each bin for the full TNG50 sample (red), as well as the GSE subsample described in Sec.~\ref{sec:gse_subsample} (orange). We specifically focus on the inner galaxy ($r<0.1~R_{\rm{dyn}}$, solid lines) and the outer galaxy ($r \in [0.1,0.5]~R_{\rm{dyn}}$, dashed lines) and the information as it pertains to computing the infall time (A parallel figure for the stellar mass is shown in Fig.~\ref{fig:feature_comparison_stellar_mass_gse}).  The general trends between both samples are consistent, although we do note an increase in the \gls{MI} of the GSE subsample for the energy and potential, which is likely due to the fact that GSE is a very massive merger that has contributed a large number of stars to the host galaxy, and therefore has a stronger influence on the overall potential of the galaxy around a specific region of infall times. 
    \label{fig:feature_comparison_gse}}
\end{figure*}

The destructive role of eccentricity is also consistent with the
asymmetry noted by \citet{2022ApJ...926..203V}: radialization is itself most
efficient for already-eccentric, massive mergers, so eccentric infall
orbits become \emph{more} eccentric over the course of the merger, 
amplifying the violent relaxation that erodes their information
content. This creates a runaway in which an initially radial merger
both deposits debris on the most destructive orbits and undergoes the
most efficient subsequent radialization---a combination that explains
why the high-eccentricity bin shows essentially complete washout
across the full mass range probed.

\subsection{MI for Galaxies with GSE-like Mergers} \label{sec:dominant_mergers}

As we have discussed, mass affects the retention of information in two competing ways: more massive mergers bring in more stars, increasing the signal-to-noise in identifying their properties, which can also increase the \gls{MI}. However, they also induce stronger violent relaxation of the host potential and sink further into the potential well via dynamical friction, where shorter orbital times accelerate phase mixing and wash out information more quickly. What we have yet to discuss is the causality of the events; more specifically, how does the order in which mergers occur, particularly the most massive ones, affect the \gls{MI} of the different mergers? For instance, if a galaxy, similarly to the Milky Way, has a very massive merger at early times, it might erase the information of all the previous mergers.

 To investigate the causal impact of mergers on information retention, we compare the \gls{MI} distribution for the full TNG50 sample to the GSE sample defined in Sec.~\ref{sec:gse_subsample}. Recall that the GSE-like mergers are selected such that they are the dominant merger in their host's assembly history: no merger that contributes a significant number of stars within $0.5~R_\mathrm{dyn}$ has a larger mass ratio than the GSE-like merger.

We show in Fig.~\ref{fig:feature_comparison_gse} the normalized \gls{MI} distributions as a function of infall time for the full sample of 98 galaxies, as well as the sample of GSE-containing galaxies. We find that the distributions are quite consistent between the two samples overall, with the exception of $E$ and $\Phi$. The other \gls{MI} distributions are comparable, though the GSE sample encodes slightly more information in $v_\phi$ and $L_z$, especially in the outer galaxy.

The energy and potential show a peak in the \gls{MI} around $\osim 7$~Gyr ago for the outer galaxy, with the \gls{MI} reaching $\osim 70\%$ of the available information in these bins. By contrast, the full sample continues to decrease at these values, with the energy retaining $\osim 30\%$ of the information in the outer galaxy at the same infall times. Although there is quite some scatter to the \gls{MI} of the GSE subset due to the smaller sample size, the increase in this infall time region is attributed to the fact that many of these galaxies have their dominant GSE merger just preceding that interval (8--10~Gyr ago), as can be seen in Fig.~\ref{fig:timing_dominant_merger}. These mergers, \emph{with 1--2~Gyr delay}, reshape the gravitational potential of the galaxy. Interestingly, the inner galaxy's \gls{MI} decreases rapidly in the 4--8~Gyr ago region. This is likely due to violent relaxation affecting the inner galaxy more than the outer galaxy, and therefore, for some (more massive) mergers, the information washout is more evident. 
Given that our selection ensured that no other merger that contributes a significant stellar population has a larger merger ratio, the imprint of the GSE is well reflected in \gls{MI} distribution by increasing the amount of information in the gravitational potential and energy in the outer galaxy, and reducing it in the inner galaxy.

\subsection{Functional Form for $\mathcal{I}(M_\star, t)$} \label{sec:functional_form}

\begin{figure*}[t] 
    \centering
    \includegraphics[width=0.95\linewidth]{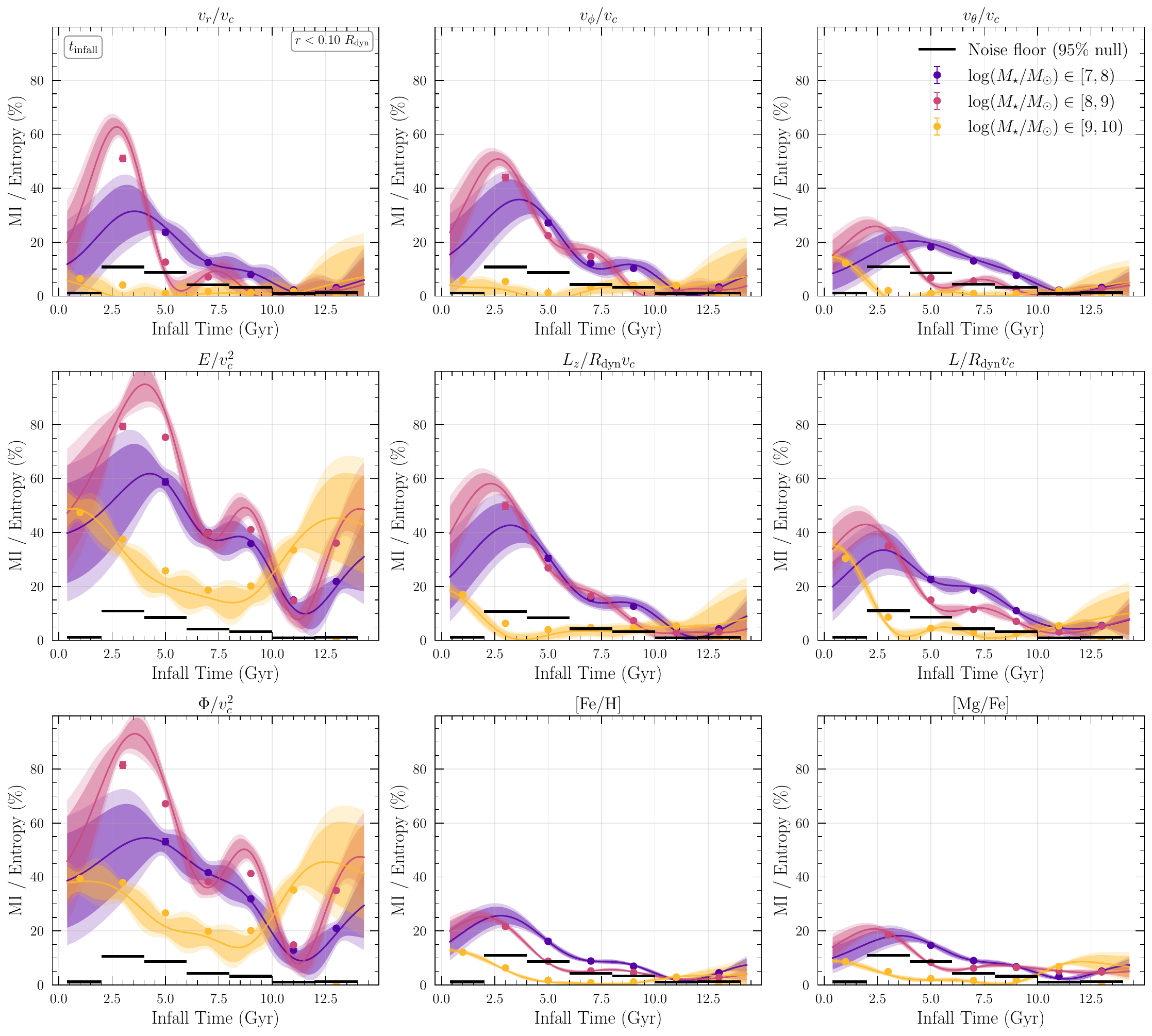}
    \caption{Fitting the Gaussian Process to the \gls{MI} data in the inner galaxy ($r < 0.1 R_{\rm{dyn}}$). See Sec.~\ref{sec:functional_form} for the fitting procedure. The inner colored intervals are those predicted by the GP, while the outer (lighter) ones are the GP errors inflated by $37\%$ to match our estimates of the errors. The black lines are the maximal noise floor at each time bin and across all mass bins. \label{fig:functional_form_fit}}
\end{figure*}

Based on the results of the analysis of 98 galaxies from TNG50, we can attempt to fit a functional form to the \gls{MI} as a function of merger stellar mass and infall time. We have tried several functional forms, including exponential decay, power-law decay, and a combination of both, and each of these has its challenges. We therefore settled on using a Gaussian process (GP) to fit the \gls{MI} as a function of stellar mass and time, which allows us to capture the complex trends in the data without imposing a specific functional form.

We model the \gls{MI} fraction $\mathcal{I}(M_\star, t)$ as a latent smooth function\footnote{Technically, $\mathcal{I}(M_\star, t)$ depends separately on the feature considered (e.g. $v_r$, $E$, etc.) as well as the target prediction (stellar mass or infall time of the merger). We will however avoid adding additional labels for simplicity of notation.} over the two-dimensional space of merger stellar mass $M_\star$ and infall time $t$, interpolated using GP regression \citep{Rasmussen2006}. The GP is constructed from data points corresponding to the mean \gls{MI} fraction within a rectangular bin $B_i = [M_{\star,\mathrm{lo}}^{(i)},\; M_{\star,\mathrm{hi}}^{(i)}] \times [t_{\mathrm{lo}}^{(i)},\; t_{\mathrm{hi}}^{(i)}]$, where $M_{\star,\mathrm{lo}}^{(i)}$ and $M_{\star,\mathrm{hi}}^{(i)}$ are the bin edges for the stellar mass and $t_{\mathrm{lo}}^{(i)}$ and $t_{\mathrm{hi}}^{(i)}$ are the bin edges for the infall time. The regression is therefore constructed as an average of the latent field rather than a point evaluation.\footnote{Treating it as a point measurement at the bin center would introduce a systematic smoothing bias and artificially prefer short correlation lengths.} We therefore employ a \textit{bin-integrated kernel} $\tilde{k}$, defined as
\begin{equation}\label{eq:GP_kernel}
\tilde{k}(B_i, B_j) = \frac{1}{|B_i||B_j|} \int_{B_i} \int_{B_j}\!\! k(\mathbf{x}, \mathbf{x}')\; d\mathbf{x}\, d\mathbf{x}',
\end{equation}
where $|B_i| = (M_{\star,\mathrm{hi}}^{(i)} - M_{\star,\mathrm{lo}}^{(i)})(t_{\mathrm{hi}}^{(i)} - t_{\mathrm{lo}}^{(i)})$ is the bin area and $k$ is the squared-exponential RBF\footnote{RBF stands for the radial basis function, adopted historically from the radial basis function networks developed in~\cite{broomhead1988radial}.} kernel~\citep{Rasmussen2006}. We have verified that the choice of the kernel does not affect the results. %
Details of the implementation and computational optimizations are provided in Appendix~\ref{sec:gp_fit}.

In Fig.~\ref{fig:functional_form_fit}, we show an example of fitting this functional form to the \gls{MI} of the normalized features and the infall time in the inner galaxy ($r < 0.1~R_{\rm{dyn}}$). The different colors show the best fit data for different mass bins for each of the features.
The GP posterior uncertainty is a lower bound on the true predictive error; Leave-One-Out calibration, in which we systematically leave out one data point at a time and refit the model to understand the error estimates, gives a standard error of $\osim 1.37$ (where this is dimensionless, it should be exactly 1 for a perfectly calibrated GP), so reported uncertainty bands should be interpreted conservatively, i.e. they need to be later inflated by $\osim 37\%$, shown in low opacity in Fig.~\ref{fig:functional_form_fit}. The fitted function captures the overall trends in the data, including the decrease in \gls{MI} with increasing infall time and the dependence on merger mass. The GP fit allows us to interpolate the \gls{MI} to infall times and masses that are not directly probed by our simulations, which is crucial for estimating the memory timescale as a function of merger mass, as we discuss in the next section.

\subsection{Diagnostics: Memory Timescale} \label{sec:memory_timescale}

We define the memory timescale $t_{\rm max}(M_\star)$ as the largest infall time $t$ for which the GP-predicted \gls{MI} meaningfully exceeds the noise floor. This helps us establish the lookback time beyond which we can no longer reliably infer merger properties from present-day dynamics as a function of merger mass.

To compute $t_{\rm max}(M_\star)$, we first have to establish the noise floor. 
The \gls{MI} estimator returns a positive value even in the absence of a true signal, due to finite sample size and estimator bias. To establish a data-grounded noise floor, we apply a permutation null test in each $(M_\star,\; t)$ bin: the target labels (merger mass, infall time, etc.) are shuffled $N_{\rm perm} = 100$ times while holding the kinematic features fixed, and the \gls{MI} fraction is recomputed for each shuffle. The 95th percentile of the resulting null distribution, $\varepsilon_{\rm null}$, represents the \gls{MI} level that would be reached by chance at the $5\%$ significance level given the actual sample size and feature distribution of that bin. The noise floor used for the full radial shell is the most conservative (maximum) value of $\varepsilon_{\rm null}$ across all bins, ensuring no detection is claimed in regimes where estimator bias is highest.

Formally, $t_{\rm max}(M_\star)$ for a specific radius is the solution to the equation
\begin{equation} \label{eq:tmax_definition}
P\left(\mathcal{I} > \varepsilon_{\rm null} \big| \mathrm{data}\right) = 1 - \mathcal{F}\left(\frac{\varepsilon_{\rm null} - \mu(M_\star, t)}{\sigma_{\rm GP}(M_\star, t)}\right) \geq P_{\rm exc},
\end{equation}
where $\mathcal{F}$ is the standard normal cumulative density function (CDF),  $\mu(M_\star,\; t)$ and $\sigma_{\rm{GP}}(M_\star,\; t)$ are the posterior mean and standard deviation of the GP fit to the binned \gls{MI} fractions \textit{at specific radii}, and $P_{\rm exc} = 0.9$ is our detection threshold. The GP posterior is Gaussian at each point, so the left-hand side is simply the survival function of the normal distribution $\mathcal{N}(\mu, \sigma^2_{\rm GP})$ evaluated at $\varepsilon_{\rm null}$, i.e. the total GP-predicted probability mass above the \gls{MI} noise floor. Uncertainty on $t_{\rm max}$ is quantified by reporting the contours at $P_{\rm exc} = 0.95$ (conservative) and $P_{\rm exc} = 0.68$ (permissive).

We will further discuss the implications of $t_{\rm max}(M_\star)$ for reconstructing the Milky Way's merger history in Sec.~\ref{sec:how_far_back}.

\section{Implications for the Milky Way} \label{sec:implications}

Having established the main drivers of information washout in a statistical sample of Milky Way--analog galaxies, we can now discuss the implications for reconstructing the merger history of our own Galaxy, particularly the GSE accretion event~\citep{2018MNRAS.478..611B,2018Natur.563...85H}. To that end, in this section we want to answer two main questions: 
\begin{enumerate}
    \item \textbf{What can the information content of the GSE debris tell us about the properties of the GSE merger itself?} Given the discovery of the GSE merger---mainly from its strong radial velocity signature in the inner halo---can we disentangle its signature from other radial events?
    \item \textbf{How far back can we see in the Milky Way's merger history given the expected information washout?} With the expected information washout from mergers, we can use our results to estimate how far back in time we can reliably reconstruct the Milky Way's merger history from present-day dynamics, and what that implies for the properties of mergers that we can hope to detect.
\end{enumerate}

\subsection{What can the information content tell us about the GSE?} \label{sec:gse_properties}

\subsubsection{Literature review of GSE properties} \label{sec:gse_literature_review}

The GSE is the most-studied merger in the Milky Way and has been characterized through a variety of observational channels. Estimates of its stellar mass converge on $\osim 2\times10^8$--$10^9~\Msun$, with an infall time of $\osim8$--10~Gyr ago. The methods used to derive these estimates fall into three broad categories: (i) \textit{dynamical methods,} using $(E, L_z)$ clustering of debris or full action-space analysis~\citep{2018MNRAS.478..611B,2018Natur.563...85H,2020ApJ...901...48N}; (ii) \textit{chemical methods,} using [Fe/H], [$\alpha$/Fe], and comparison to chemical evolution models~\citep{Mackereth2018,Feuillet2020,Feuillet_2021_GSE_ages}; and (iii) \textit{age-based methods,} using CMD fitting, isochrone fitting, or asteroseismology of individual stars~\citep{2020ApJ...897L..18B,Montalban2021,age_GSE_I,Woody2025}. Table~\ref{tab:literature} summarizes these estimates, and additionally includes estimates for the properties of other Milky Way mergers.

\begin{table}
\centering
\begin{tabular}{lccc}
\hline\hline
Progenitor & $t_{\rm infall}$ [Gyr ago] & $\log_{10}(M_\star/\Msun)$ & Source \\
\hline
GSE & $9.1^{+0.7}_{-0.7}$ & $8.43^{+0.15}_{-0.16}$ & K20 \\
    & $10.5$ & $8.7$ & N22 \\
    & $9.5^{+0.8}_{-1.4}$ & $9.0^{+0.3}_{-0.3}$ & S26 \\
\hline
Helmi streams & $5$--$8$ & $\osim8$ & Ko19 \\
              & $10.1^{+0.7}_{-1.2}$ & $7.96^{+0.19}_{-0.18}$ & K20 \\
              & $7.9$ & $8$ & N22 \\
              & $10.1^{+0.7}_{-0.9}$ & $8.4^{+0.2}_{-0.2}$ & S26 \\
\hline
Heracles & $10.5$--$11.6$ & $\osim8.7$ & H21 \\
         & $10.5^{+0.5}_{-0.7}$ & $8.2^{+0.2}_{-0.2}$ & S26 \\
\hline
I'itoi & $12.3^{+0.6}_{-0.7}$ & $6.7^{+0.4}_{-0.4}$ & S26 \\
\hline
LMS-1/Wukong & $\osim8$ & $6$--$7$ & Ma21 \\
             & $8.3$ & $7.1$ & N22 \\
             & $12.6^{+0.4}_{-0.5}$ & $7.3^{+0.4}_{-0.3}$ & S26 \\
\hline
Sagittarius & $6.8^{+1.1}_{-1.1}$ & $8.44^{+0.22}_{-0.21}$ & K20 \\
            & $5.9$ & $8.8$ & N22 \\
            & $7.2^{+1.1}_{-1.2}$ & $8.8^{+0.2}_{-0.2}$ & S26 \\
\hline
Sequoia (K19) & $11.3^{+0.6}_{-0.6}$ & $7.4^{+0.5}_{-0.4}$ & S26 \\
Sequoia (M19) & $10.3^{+0.7}_{-0.9}$ & $8.5^{+0.2}_{-0.3}$ & S26 \\
Sequoia (N20) & $11.4^{+0.6}_{-0.7}$ & $7.3^{+0.5}_{-0.4}$ & S26 \\
Sequoia & $9.4^{+0.4}_{-0.5}$ & $7.9^{+0.1}_{-0.1}$ & K20 \\
        & $\osim9$ & $\osim7.7$ & My19 \\
\hline
Thamnos & $11.8^{+0.5}_{-0.6}$ & $8.1^{+0.4}_{-0.4}$ & S26 \\
\hline\hline
\end{tabular}
\caption{Literature estimates of infall time and stellar mass for Milky Way merger progenitors, plotted in Fig.~\ref{fig:tmax_infall_time_mass}. Approximate values (prefixed with $\sim$ in the original reference) are shown with gray markers in the figures. Source abbreviations: K20 = \citet{2020MNRAS.498.2472K}, V19 = \citet{689481513cbfd9e89b25aab7aea0d35f93b1f06e}, N22 = \citet{2022arXiv220409057N}, S26 = \citet{Sante2026}, Ko19 = \citet{3f3327b6a6c420f1f8f0e755c20514f75df9e46e}, H21 = \citet{2021MNRAS.500.1385H}, Ma21 = \citet{2021ApJ...920...51M}, My19 = \citet{a0487e1ec9180009817123819c555c308e51c2c3}. Sequoia has multiple names because \cite{Sante2026} use different selections to compute the properties of the same merger.}
\label{tab:literature}
\end{table}

\subsubsection{The radial velocity signature} \label{sec:radial_velocity_puzzle}
\begin{figure*}[t]
    \centering
    \includegraphics[width=0.95\linewidth]{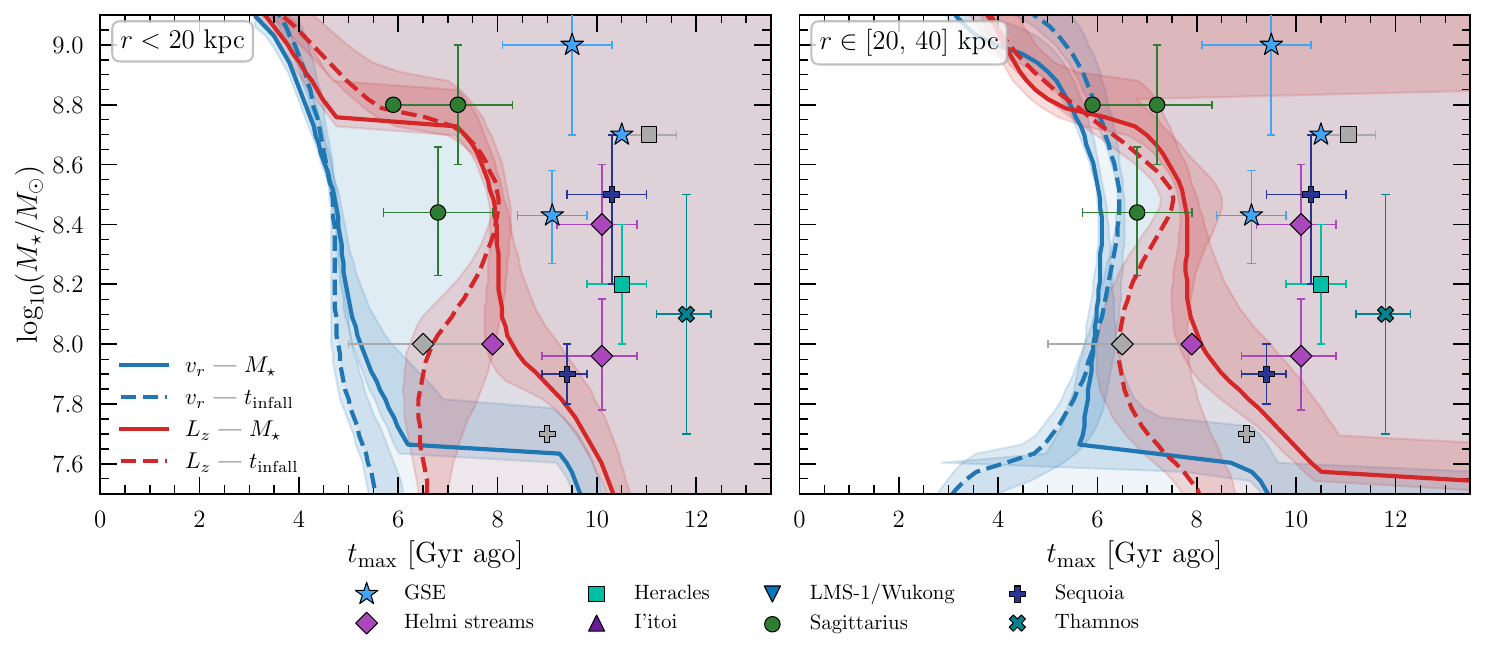}
    \caption{The stellar mass of the progenitor as a function of the maximum infall time $t_{\rm{max}}$ for which the \gls{MI} in the radial velocity ($v_r$, in blue) and angular momentum ($L_z$, in red) is significantly different from zero. The shaded region is where we do not expect enough information based on the stellar mass or the infall time to deduce the merger history. 
    Added are the literature estimates for the infall times and stellar masses of various past accretion events as summarized in \cite{Sante2026} and detailed in Tab.~\ref{tab:literature}. 
    The gray markers correspond to estimates that are approximate in the original reference, while the colored markers correspond to more confident estimates, mostly  with reported uncertainties. Note that the stellar mass estimates for I'itoi and LMS-1/Wukong are too small for the corresponding points to appear in this figure.
    \label{fig:tmax_infall_time_mass}}
\end{figure*}

A hallmark of the GSE discovery was its striking signature in radial velocity
space---the ``Sausage'' of highly radial orbits~\citep{2018MNRAS.478..611B}.
However, our information-theoretic analysis frames this observation in a new light.
As seen in Fig.~\ref{fig:functional_form_fit}, the \gls{MI} carried by
energy $E$ and potential $\Phi$ remains above zero consistently across all
masses and infall times, while
for other features, the \gls{MI} drops to being consistent with zero at infall times of $\sim$5--6\,Gyr ago
for the most massive mergers and $\sim$8--10\,Gyr ago for the least massive
ones. This suggests that we can expect to retain \emph{some} information about
mergers that occurred up to $\sim$8--10\,Gyr ago in the non-energy features, particularly if they were not
too massive; for more massive mergers, the information washout is more severe,
limiting our view to roughly the last 5--6\,Gyr.

More specifically, Fig.~\ref{fig:functional_form_fit} shows that the \gls{MI}
carried by the radial velocity $v_r$ in the inner galaxy ($r < 0.1~R_{\rm dyn}$)
is consistent with zero for older infall times. \emph{This does not mean that the GSE is younger than previously expected, but rather that its radial velocity signature cannot be reliably used to infer its properties if it is older than $\osim$5--6\,Gyr, according to the typical information retention timescales we measure in TNG50.} This is analogous to dropping ink into water: the color is visible long after the molecules have become indistinguishable from the background. The $v_r$ signature demonstrates that a merger occurred, but does not, on its own, diagnose its properties. %

In Fig.~\ref{fig:tmax_infall_time_mass}, we apply this information-theoretic perspective and quantify
the information washout process by showing the stellar mass of the progenitor as a function of the maximum
infall time $t_{\rm max}$, defined in Sec.~\ref{sec:memory_timescale}, for which the \gls{MI} in $v_r$ and angular momentum $L_z$ are
significantly different from zero. To define this threshold, we shuffle the
infall-time labels 100 times and compute the \gls{MI} for each permutation, yielding a
null distribution; we reject the null hypothesis if the observed \gls{MI} exceeds the
90th percentile of this distribution as explained above.

The blue and red curves in Fig.~\ref{fig:tmax_infall_time_mass} trace the
$M_\star$--$t_{\rm max}$ relationship for $v_r$ and $L_z$ in the inner galaxy
($r < 0.1~R_{\rm dyn}$) (left) and slightly larger radii ($r\in[0.1,0.2]~R_{\rm dyn}$) (right), respectively; the shaded region marks the regime in which we do not
expect to retain sufficient information to recover the merger's properties. To scale the radii to the Milky Way, we use $R_{\rm dyn} = 200$~kpc, which is consistent with the modeling of~\cite{2025arXiv250305877O}. The GSE's debris is predominantly found at radii $\lesssim25$~kpc, with measurements ranging from the solar neighborhood~\citep[see, e.g.,][]{2018MNRAS.478..611B,2018Natur.563...85H,2019ApJ...874....3N} and extending into the stellar halo~\citep{Naidu2021}. Given this spatial extent, the left panel is more relevant for the GSE.
Overplotted with star markers
are literature estimates for the GSE, 
as summarized in \cite{Sante2026} and in Table~\ref{tab:literature}. 
Strikingly, these estimates place the GSE squarely in or near the shaded region,
suggesting that the radial velocity signature of the GSE is not informative about its properties. 

\subsubsection{Which features carry information about the GSE?} \label{sec:resolutions_radial_velocity_puzzle}

\begin{figure*}
    \centering
    \includegraphics[width=0.70\linewidth]{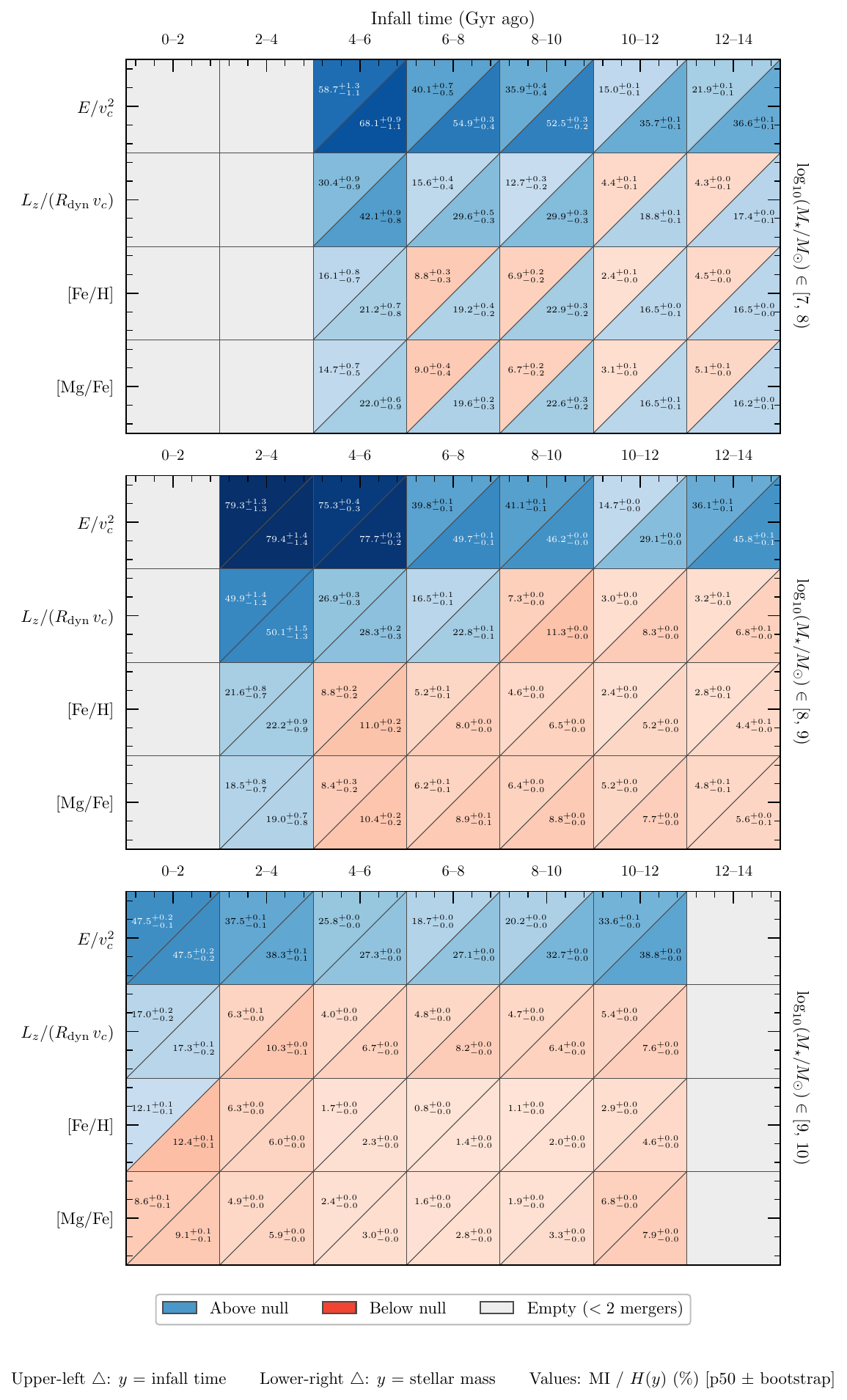}
    \caption{We isolate the \gls{MI} of the four variables ($E, L_z, \mathrm{[Fe/H]}, \mathrm{[Mg/Fe]}$) as a function of infall time (columns of each grid) for three stellar mass bins (the three grids). Here we are only considering the inner galaxy ($r < 0.1~R_{\rm{dyn}}$). For each quadrant, the top left triangle shows the \gls{MI} for inferring the infall time, while the bottom right triangle shows the \gls{MI} for inferring the stellar mass.
    The bins are colored by whether they are higher than the \emph{maximal null value $\varepsilon$} for the \gls{MI} between that specific feature and the target variable (infall time or stellar mass) across all bins, which is a conservative threshold for claiming a detection. Blue regions are above this threshold, while red regions are below it.
    The gray region indicates we only have a single merger, and therefore we cannot meaningfully compute the \gls{MI}.
    \label{fig:information_retention_old_mergers}}
\end{figure*}

Fig.~\ref{fig:information_retention_old_mergers} isolates the four features most commonly used in dynamical and chemo-dynamical inference---$E$, $L_z$, [Fe/H], and [Mg/Fe]---and shows, for the inner galaxy  ($r < 0.1~R_{\rm dyn}$), their normalized \gls{MI} in bins of infall time and progenitor stellar mass. Each bin is colored by whether the measured \gls{MI} exceeds a permutation null threshold: blue bins carry statistically significant information, red bins do not. Here, for consistency across bins, we choose the highest noise floor in that row (across infall times). 

For mergers in the GSE-like bin ($\log M_\star/\Msun \in [8, 9]$, $t_{\rm infall} \sim 8$--10~Gyr ago), we read off the following from Fig.~\ref{fig:information_retention_old_mergers}: $E$ carries $\sim$40--50\% of the available information, remaining comfortably above the null, while $L_z$, [Fe/H], and [Mg/Fe] carry only $\sim$5--10\%, which does not exceed the threshold for significance. The information budget for this region of parameter space is therefore dominated by $E$, with the other three features contributing sub-dominant or null amounts.

This decomposition has a direct implication for previous characterizations of the GSE in the literature. Methods relying on $v_r$ or $L_z$ alone (e.g. the original ``Sausage'' identification of \citealt{2018MNRAS.478..611B} with radial velocity) should be interpreted as detection signals: they indicate that a merger occurred, not what its properties were. Methods relying on $E$---whether manual $(E, L_z)$ clustering~\citep{2018Natur.563...85H,2020ApJ...901...48N}, unsupervised machine learning~\citep{2020ApJ...903...25N,2022MNRAS.509.5992S}, or simulation-based inference over $(E, L_z, \mathrm{[Fe/H]}, \mathrm{[Mg/Fe]})$~\citep{Sante2026}---are effectively $E$-dominated in this region of parameter space.

Finally, we note that our framework does not address age-based methods~\citep{2020ApJ...897L..18B,Woody2025,age_GSE_I}, because stellar ages are set at formation and are not subject to dynamical mixing. Ages should in principle retain full information about progenitor properties indefinitely, limited only by measurement precision. However, stellar ages are challenging to measure accurately, which is why they are not included in most analyses, and therefore we do not consider them here. %

We caution that the chemical \gls{MI} values in Fig.~\ref{fig:information_retention_old_mergers} are derived from the IllustrisTNG chemical enrichment model, which has known limitations~\citep{2018MNRAS.477.1206N,2018MNRAS.473.4077P}. The real Milky Way may carry more chemical information than TNG50 captures; this can already be seen by the cross check with the FIRE-2 galaxies of Fig.~\ref{fig:feature_comparison}, in which [Fe/H] and [Mg/Fe] retain more information in the FIRE-2 simulations than TNG50. Our result that $E$ dominates the $(E, L_z, \mathrm{[Fe/H]}, \mathrm{[Mg/Fe]})$ information budget should therefore be read as a lower bound on chemistry's contribution in the real Galaxy.

\subsection{Implications for reconstructing the merger history of Milky Way} \label{sec:how_far_back}

Beyond the question of \emph{whether} information is retained, we can ask
\emph{where the horizon of information} in the galaxy is. To address this point, we show in Fig.~\ref{fig:tmax_infall_time_mass}, 
the maximum time $t_{\rm max}$ at which the \gls{MI}
for inferring the merger's infall time and stellar mass is significantly
different from zero, for $r < 20$~kpc and $r \in [20, 40]$~kpc. 

We focus on results for $v_r$ and $L_z$, and do not analyze the information content of the orbital energy, as it retains significant information for all masses and infall times at all galactocentric radii (see Figs.~\ref{fig:feature_mass_comparison} and \ref{fig:information_retention_old_mergers}). 

\paragraph{Angular
momentum $L_z$} for the two radial bins considered ($r<20$~kpc and $r\in[20, 40]$~kpc), information about the infall time is retained for $\lesssim 6$--8~Gyr, but it
is highly degraded for massive mergers, with $t_\mathrm{max}$ decreasing to $4$~Gyr for mergers with stellar mass of $10^9~\Msun$. 
$L_z$ retains information about the stellar mass for similar timescales, though it is washed out $\osim2$~Gyr sooner for the lower mass mergers. 
Beyond $\osim 8$~Gyr, merger information is fully washed out from
the angular momentum, making $L_z$ a poor tracer of the ancient merger history
of the Milky Way. 

\paragraph{Radial velocity $v_r$} For a detailed discussion of $v_r$, see Sec.~\ref{sec:radial_velocity_puzzle}.
The timescale on which $v_r$ retains information is comparable to that for the angular momentum in the $r\in [20, 40]$~kpc bin, but these noticeably diverge in the inner galaxy, where $v_r$ retains much less information. This is expected, as $L_z$ is almost conserved for axisymmetric potentials, while $v_r$ is not a conserved quantity and is more sensitive to the details of the merger and the host potential. This makes $v_r$ a poor tracer of the merger history of the Milky Way beyond 5--6~Gyr ago, particularly in the inner galaxy.

This is important to understand both for a more robust interpretation of existing merger history inference techniques and also to guide future efforts. Most of the inferred stellar masses and infall times in the literature shown in Fig.~\ref{fig:tmax_infall_time_mass} are in the regime where we do not expect to retain information about the merger properties, which suggests that their identification is likely driven by $E$ rather than $L_z$ or kinematics. 

Various methods have been implemented to build the merger tree of the Milky Way, and our information-theoretic framework allows us to reinterpret the constraining power of each of the three broad classes below:

\paragraph{Manual selection of overdensities} Selection by eye in $(E,\; L_z)$ or $(E,\; L_z,\; \mathrm{chemistry})$ space~\citep[][among others]{1999Natur.402...53H,2002ApJ...569..245N,2020ApJ...901...48N,2022arXiv220404233H,2021ApJ...919...66B} is reliable when the selection is $E$-driven. Selection criteria that rely primarily on $L_z$ thresholds should be interpreted with caution for mergers older than $\sim$6\,Gyr, where Fig.~\ref{fig:tmax_infall_time_mass} indicates that $L_z$ carries little information about merger properties. However, the signatures might still be informative for detection.

\paragraph{Unsupervised machine learning} Clustering algorithms applied to $(E,\; L_z)$ or extended feature sets~\citep{2020ApJ...903...25N,2020NatAs...4.1078N,2020MNRAS.498.2472K,2022MNRAS.510.2407B,2022ApJ...926...26S} identify structures based on density in feature space. Our framework predicts that such algorithms will find structures primarily via their $E$-signatures, with $L_z$ and chemistry providing secondary discriminating power.

\paragraph{Simulation-based inference} The recent application of normalizing flows to $(E,\; L_z,\; \mathrm{[Fe/H]},\; \mathrm{[Mg/Fe]})$ by \citet{Sante2026} represents the current state of the art for data-driven Milky Way assembly reconstruction. Fig.~\ref{fig:information_retention_old_mergers} implies that in the GSE-mass regime, their posterior is effectively dominated by the $E$ channel, with $L_z$ contributing a sub-dominant correction and chemistry providing a small additional signal. The precision of their reported posteriors should be interpreted in light of this decomposition: it reflects the constraining power of $E$ more than of the joint feature set.

\subsection{Limitations}

Throughout this work, we have characterized the information content of each observable individually. 
A study that uses many observables simultaneously, however, will in general recover a different information content:
when features carry overlapping, redundant information, the \gls{MI} contained by the set is smaller than the sum of the individual \gls{MI}s;
alternatively, features may combine synergistically, with more information contained in the set than would be expected from the sum of the individual parts.%
\footnote{For example, consider a random variable $Y$ that is the sum of independent random variables $X_1$ and $X_2$. Knowing just $X_1$ or $X_2$ individually is insufficient to determine $Y$, though the values will be correlated and each $X_i$ contains some information on $Y$. Knowing both $X_1$ and $X_2$, however, completely determines $Y$; these features sum synergistically.}
Among our kinematic observables, substantial redundancy is expected, since $E$, $\Phi$, and the velocity components are deterministically related, and $L \approx L_z$ in the near-spherical halo limit. %
The more interesting case is the combination of kinematics and chemistry: for old mergers, where kinematic \gls{MI} has decayed and the chemistry floor is low but non-zero, the sum of the individual \gls{MI} are well below the maximum 100\% value and synergistic gains are physically plausible. %
For example, two progenitors with similar present-day phase-space distributions but distinct chemical-evolution histories would be jointly distinguishable while remaining individually ambiguous. %
The per-observable \gls{MI} curves presented here therefore should be read as a feature-by-feature diagnostic rather than as an additive accounting of total recoverable information; quantifying the redundant and synergistic contributions through a partial information decomposition~\citep{Williams2010}, and propagating them into joint detection horizons, is a natural extension that we leave for future work. %
Such a decomposition would clarify the structure of the joint information implicitly captured by joint-density approaches such as the normalizing-flow analysis of~\citet{Sante2026}. %

\section{Conclusions}
\label{sec:conclusions}

We have quantified the information washout of the merger history of Milky Way--like galaxies using \gls{MI} applied to 98 analogs from the TNG50 simulation. By measuring how much information about merger properties---infall time and stellar mass---is encoded in the present-day chemodynamics of accreted stars, we identify the key physical processes that control information retention and map out the observational horizon for reconstructing the Galaxy's assembly history. Our main conclusions are as follows:

\begin{enumerate}

\item \textbf{Energy and gravitational potential are the most informative and longest-lived tracers.} The normalized energy $E$ and potential $\Phi$ retain the most \gls{MI} about merger properties across all infall times and radial bins, with median normalized \gls{MI} of $\osim 0.3$ across the full galaxy sample (Fig.~\ref{fig:feature_importance}). They remain informative even for mergers that occurred $\gtrsim 10$~Gyr ago, making them the most robust dynamical tracers of the merger history. This makes it critical to keep developing robust methods to accurately obtain the gravitational potential of the Milky Way.

\item \textbf{Radial velocity information is short-lived.} The \gls{MI} carried by $v_r$ decays rapidly and is consistent with zero for mergers older than $\osim5$--6~Gyr in the inner galaxy (Fig.~\ref{fig:feature_comparison}). This is driven by stochastic barycenter displacements during mergers and the causal dependence of $v_r$ evolution on prior merger history, both of which render the mapping from merger parameters to present-day $v_r$ many-to-one across the galaxy population. Azimuthal quantities like $L_z$ and $v_\phi$ decay more slowly, reflecting the approximate axisymmetry of the host potentials.

\item \textbf{Chemical abundances are immune to dynamical mixing but carry limited information.} The \gls{MI} of [Fe/H] and [Mg/Fe] show little dependence on infall time or radius, confirming their role as intrinsic stellar properties unaffected by phase mixing (Fig.~\ref{fig:radial_split}). However, their overall \gls{MI} is lower than that of the dynamical features, likely due to overlapping abundance patterns between different mergers in the IllustrisTNG chemical enrichment model. Checking these results on three FIRE-2 Milky Way--like galaxies shows similar patterns, albeit with the FIRE galaxies chemical abundances holding onto more information than TNG50 (Fig.~\ref{fig:feature_comparison}). Chemistry still remains valuable as complementary tracers, particularly when dynamical information has been erased.

\item \textbf{Information washout is controlled by three factors: radial position, infall time, and merger mass.} Stars in the inner galaxy ($r < 0.1~R_{\rm{dyn}}$) lose information faster due to shorter orbital times and more rapid phase mixing. The number of completed orbits $N_{\rm{orb}} \approx t_{\rm{infall}}/\tau_{\rm{dyn}}(r)$ provides a first-order framework for predicting information retention (Fig.~\ref{fig:infall_radius_2d}). The most massive mergers retain the least information despite contributing the most stars, as violent relaxation and dynamical friction---which drives them to the bottom of the potential well where orbital times are shortest---compound to erase their dynamical signatures on short timescales (Fig.~\ref{fig:feature_mass_comparison}).

\item \textbf{Orbital eccentricity sets a ceiling on information retention.} Mergers on highly radial orbits ($e \gtrsim 0.6$) show uniformly low \gls{MI} regardless of stellar mass, as pericentric shocks and violent relaxation erase their phase-space signatures (Fig.~\ref{fig:orbital_params_mass_split}). At low eccentricities, a strong mass dependence emerges, with more massive mergers retaining more information due to their larger debris counts and wider energy distributions.

\item \textbf{The dominant merger leaves a localized imprint on the \gls{MI} budget.} We compared the \gls{MI} as a function of infall time for the GSE-like subsample with the full TNG50 sample (Sec.~\ref{sec:dominant_mergers}) and found that the overall trends are consistent between the two. The most notable difference, however, is an enhancement in the \gls{MI} of $E$ and $\Phi$ around 6--8~Gyr ago, coinciding with the infall times of most GSE-like mergers (Fig.~\ref{fig:feature_comparison_gse}). This indicates that the gravitational potential has incorporated the imprint of this massive merger and, in the absence of subsequent comparably massive events, retains that signature to the present day.

\item \textbf{The GSE's radial velocity signature is a detection channel, not a characterization channel.} For a merger at the GSE's estimated mass and infall time, the \gls{MI} in $v_r$ is consistent with zero (Fig.~\ref{fig:tmax_infall_time_mass}). The distinctive $v_r$ signature that originally identified the GSE therefore demonstrates that a merger occurred, but cannot be used to infer its mass or infall time. Characterization of the GSE must instead rely on features that retain information at its $(M_\star,\; t_{\rm{infall}})$, primarily energy and—to a lesser extent, given the limitations of IllustrisTNG chemistry—chemical abundances (Fig.~\ref{fig:information_retention_old_mergers}).

\item \textbf{The outer halo is the most valuable region for reconstructing the ancient merger history.} For energy, information is retained at all infall times, though the information retained increases as a function of radius (Fig.~\ref{fig:radial_split}). For $L_z$ and kinematics, information increases with radius and infall time.
This makes deep, wide-field surveys of the outer stellar halo essential for constraining the early assembly history of the Milky Way.

\end{enumerate}

These results provide an information-theoretic framework for interpreting the outputs of current and future spectroscopic surveys aimed at reconstructing the Milky Way's merger history. They highlight that energy---and hence accurate distances, radial velocities, and modeling of the gravitational potential---is the single most important observable for merger reconstruction, while chemical abundances provide a complementary, time-independent channel. Methods that rely heavily on angular momentum or radial velocity should be interpreted with caution for mergers older than $\osim5$--6~Gyr, as the information content of these features is expected to be largely erased by dynamical mixing.

\begin{acknowledgments}

LN would like to particularly thank Paul Schechter for the insightful and inspiring discussions about this work. 

We also thank  H. Dominguez Sanchez, A. Harrow, A. Hussein, M. Lisanti, P. Melchior, Z. Mezghanni, X. Ou, T. Smidt, X. Zhang.

LN and AT received support
from NSF, via CAREER award AST-2337864. L.N. is also supported by the Sloan Fellowship. 
DF is supported by the Department of Energy (DOE) under Award Number DE-SC0007968 and by a grant from the Simons Foundation (MPS-SIP-00929613, ML).
NS is supported by the Brinson Foundation through a Brinson Prize Fellowship grant. %

This work used Bridges-2 at Pittsburgh Supercomputing Center through allocation phy210068p from the Advanced Cyberinfrastructure Coordination Ecosystem: Services \& Support (ACCESS) program, which is supported by National Science Foundation grants \#2138259, \#2138286, \#2138307, \#2137603, and \#2138296.

The authors acknowledge the Texas Advanced Computing Center (TACC) at The University of Texas at Austin for providing computational resources through Stampede-3 that have contributed to the research results reported within this paper. \url{http://www.tacc.utexas.edu}.

FIRE-2 simulations are publicly available \citep{Wetzel2023, 2025arXiv250806608W} at \url{http://flathub.flatironinstitute.org/fire}.
Additional FIRE simulation data is available at \url{https://fire.northwestern.edu/data}.
A public version of the \textsc{Gizmo} code is available at \url{http://www.tapir.caltech.edu/~phopkins/Site/GIZMO.html}.
The FIRE-2 simulations were generated using: XSEDE, supported by NSF grant ACI-1548562; Blue Waters, supported by the NSF; Frontera allocations AST21010 and AST20016, supported by the NSF and TACC; Pleiades, via the NASA HEC program through the NAS Division at Ames Research Center. 

The IllustrisTNG simulations were undertaken with compute time awarded by the Gauss Centre for Supercomputing (GCS) under GCS Large-Scale Projects GCS-ILLU and GCS-DWAR on the GCS share of the supercomputer Hazel Hen at the High Performance Computing Center Stuttgart (HLRS), as well as on the machines of the Max Planck Computing and Data Facility (MPCDF) in Garching, Germany.

This work made use of Anthropic's Claude (Sonnet and Opus models) (\url{https://www.anthropic.com/claude}) as an assistive tool for code development, manuscript editing, suggesting relevant literature, and discussion of conceptual aspects of the analysis. All suggested references were checked against the original sources by the authors, and all scientific content, analysis choices, and conclusions are the authors' own.

This report was prepared as an account of work sponsored by an agency of the United States Government. Neither the United States Government nor any agency thereof, nor any of their employees, makes any warranty, express or implied, or assumes any legal liability or responsibility for the accuracy, completeness, or usefulness of any information, apparatus, product, or process disclosed, or represents that its use would not infringe privately owned rights. Reference herein to any specific commercial product, process, or service by trade name, trademark, manufacturer, or otherwise does not necessarily constitute or imply its endorsement, recommendation, or favoring by the United States Government or any agency thereof. The views and opinions of authors expressed herein do not necessarily state or reflect those of the United States Government or any agency thereof. Further, any opinions, findings, and conclusions or recommendations expressed in this material are those of the authors and do not necessarily reflect the views of the National Science Foundation.

\end{acknowledgments}

\begin{contribution}

LN conceived of the project, developed the information-theoretic framework, performed the analysis, produced the figures, and wrote the manuscript. DF reconstructed the merger histories of the Milky Way--like galaxies in TNG50 and provided the accreted-star catalogs used in this work. AT performed the analogous merger-history reconstruction for the FIRE-2 galaxies. EYD and NS contributed insightful discussions and detailed comments, particularly on Sections 4 and 5. All authors read the manuscript and contributed to its final form.


\end{contribution}

%
\facilities{Bridges-2~\citep{sanielevici2021bridges}, Stampede-3}

\software{Matplotlib~\citep{Hunter:2007}, 
Numpy~\citep{harris2020array},
OverCite \citep{Shariat2026},  
          Scipy.stats~\citep{2020SciPy-NMeth}, 
Sklearn~\citep{pedregosa2018scikitlearnmachinelearningpython}
          }


\newpage
\pagebreak
\newpage

\appendix

\renewcommand{\thefigure}{\thesection\arabic{figure}}
\renewcommand{\thetable}{\thesection\arabic{table}}
\makeatletter
\@addtoreset{figure}{section}
\@addtoreset{table}{section}
\makeatother

\section{Additional Figures}

\begin{figure}[h]
    \centering
    \includegraphics[width=0.95\textwidth]{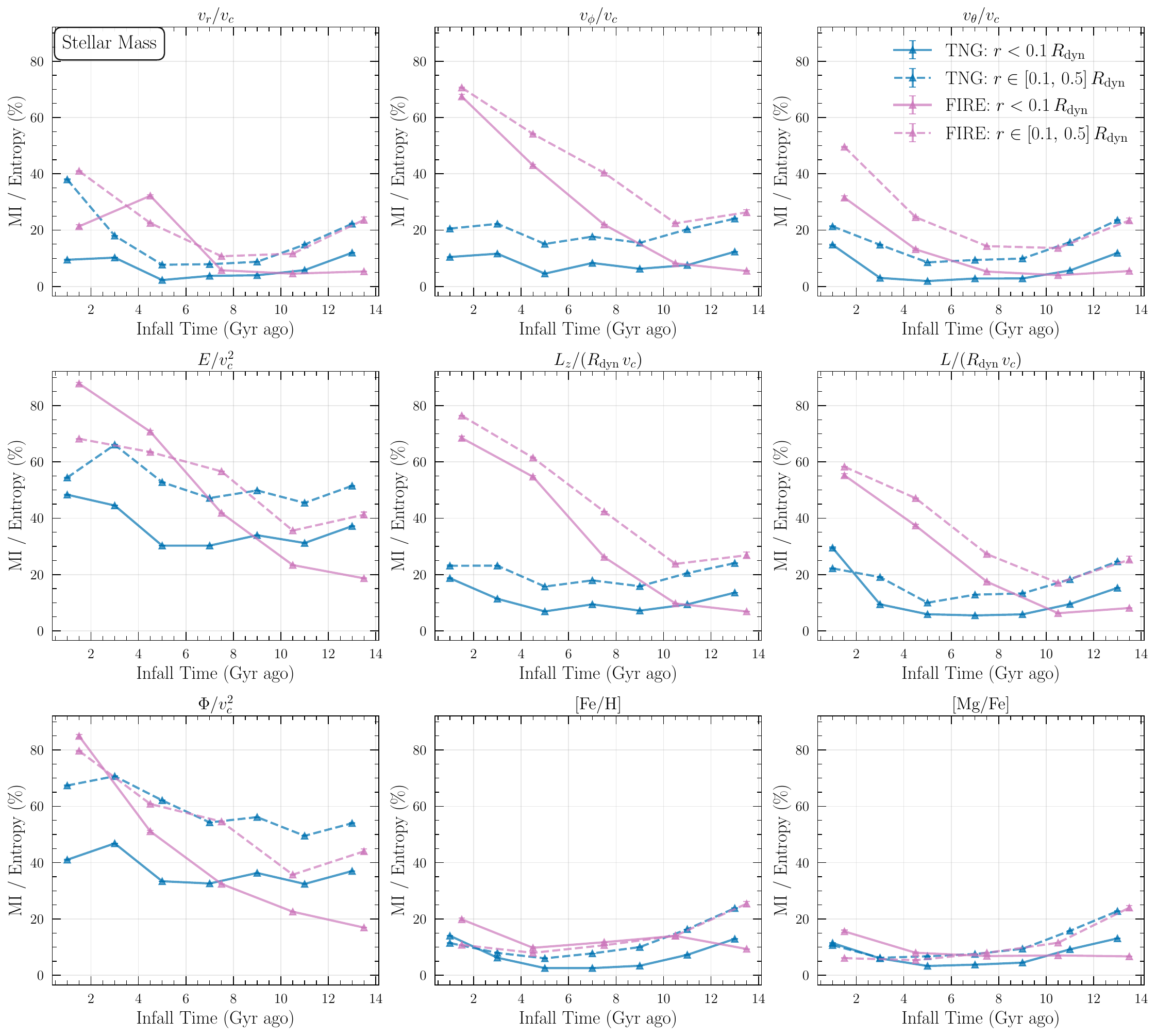}
    \caption{Similar to Fig.~\ref{fig:feature_comparison} but computing the \gls{MI} for the stellar mass instead of the infall time.
    \label{fig:feature_comparison_stellar_mass}}
\end{figure}

\begin{figure}[h]
    \centering
    \includegraphics[width=0.95\textwidth]{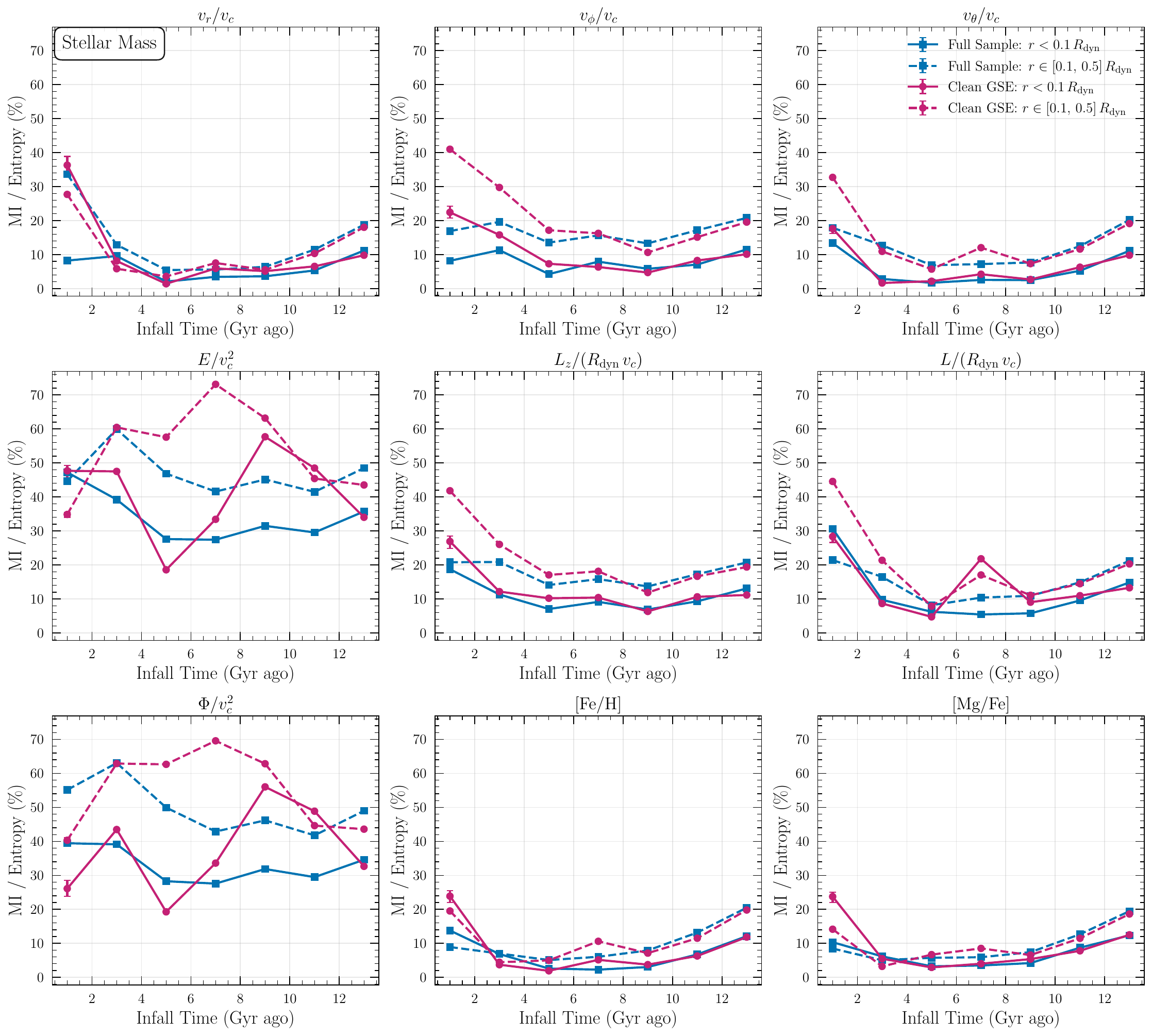}
    \caption{Similar to Fig.~\ref{fig:feature_comparison_gse} but computing the \gls{MI} for the stellar mass instead of the infall time.
    \label{fig:feature_comparison_stellar_mass_gse}}
\end{figure}

\begin{figure*}[h]
    \centering
    \includegraphics[width=0.95\linewidth]{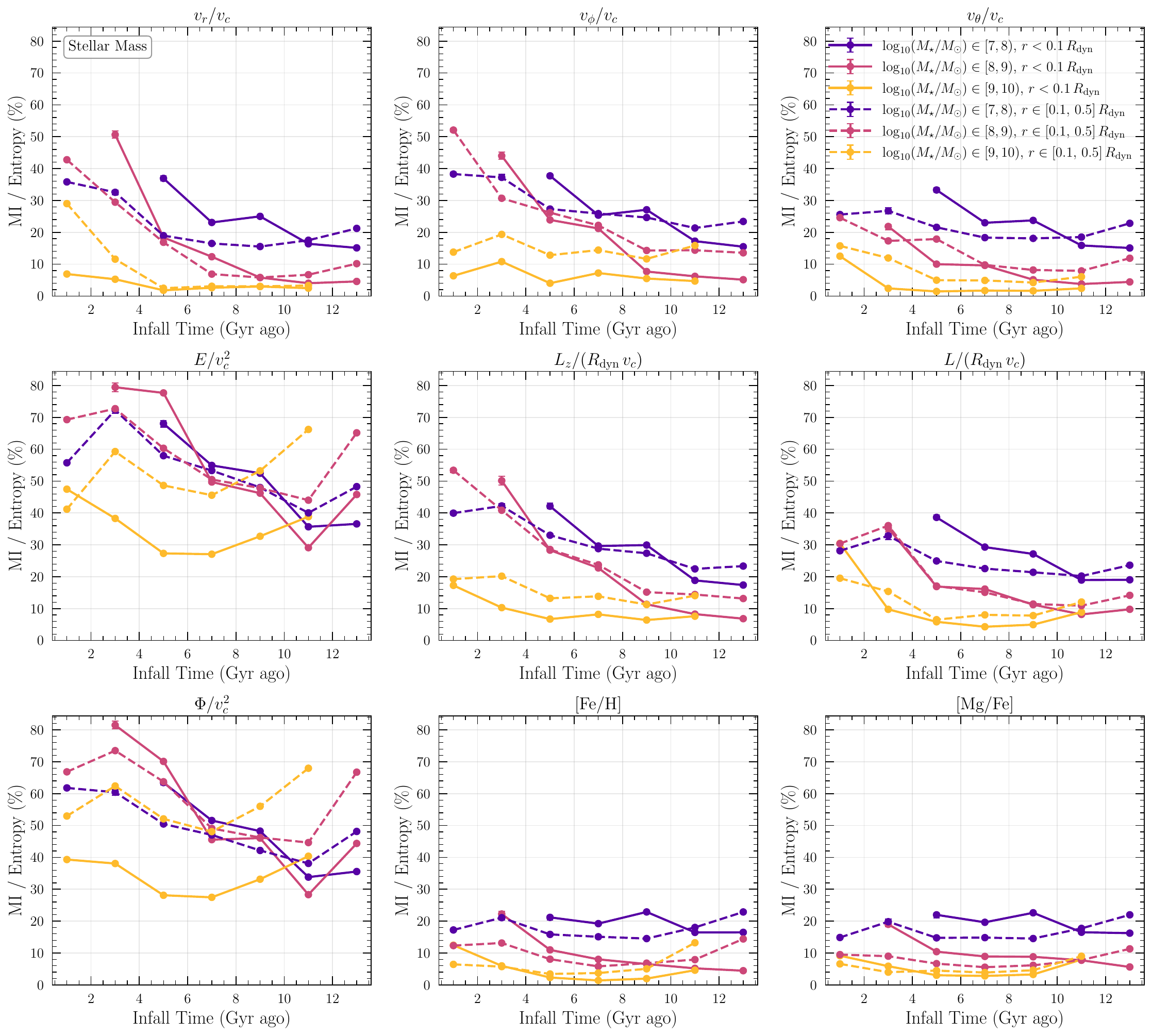}
    \caption{Similar to Fig.~\ref{fig:feature_comparison} but computing the \gls{MI} for the stellar mass instead of the infall time.  
    \label{fig:feature_mass_comparison_stellar_mass}}
\end{figure*}

\begin{figure}[h]
    \centering
    \includegraphics[width=0.45\textwidth]{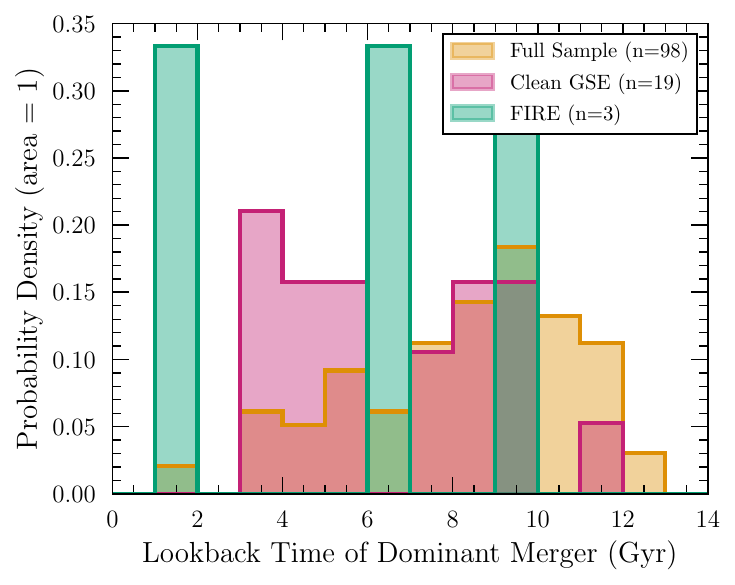}
    \caption{Histogram of the dominant merger lookback times for the Milky Way--analog galaxies in the full sample (yellow), the GSE-like sample (purple), and the FIRE sample (green).
    \label{fig:timing_dominant_merger}}
\end{figure}

In this appendix, we show additional figures that complement the main text. In particular, we show the \gls{MI} for the stellar mass instead of the infall time in Figs.~\ref{fig:feature_comparison_stellar_mass}, \ref{fig:feature_comparison_stellar_mass_gse}, and \ref{fig:feature_mass_comparison_stellar_mass}. We also show the histogram of the dominant merger lookback times for the Milky Way--analog galaxies in the full sample, the GSE-like sample, and the FIRE sample in Fig.~\ref{fig:timing_dominant_merger}.

\section{Computation of the Gaussian Process fit} \label{sec:gp_fit}

The simplest choice of computation of the GP is to evaluate it at each point in $(M_*^{(i)}, t^{(i)})$. However, this approach would not be correct, as the data itself is binned in stellar and infall times, and therefore this averaging process needs to be taken into account. To address that, the double integral from Eq.~\ref{eq:GP_kernel} is evaluated over the \emph{whole bin}. Doing so in technically challenging, which is why this is numerically using a $4\times4$ Gauss--Legendre quadrature rule. 

In the simple case of one dimension, Gauss-Legendre quadrature rule is
\begin{equation}
    \int_{-1}^{1} f(x) dx \approx \sum_{i=1}^n \omega_i f(x_i), 
\end{equation}
where $\omega_i$ are weights, and $x_i$ are roots of the nth Legendre polynomial. To generalize the limits of integration from [-1,1] to $[a,b]$, we would need to rescale the variable integrated over. 

Now, moving on two dimensions as discussed in the text and particularly Eq.~\ref{eq:GP_kernel}, for bin $B_i$, the quadrature nodes and weights are constructed by a tensor product of the one-dimensional rule over $[M_{\star,\mathrm{lo}}^{(i)},\; M_{\star,\mathrm{hi}}^{(i)}]$ and $[t_{\mathrm{lo}}^{(i)},\; t_{\mathrm{hi}}^{(i)}]$:

\begin{eqnarray}
\mathbf{x}_{ip} =& \begin{pmatrix} \bar{M}_i + h_{M,i} \xi_\alpha, ~ \bar{t}_i + h_{t,i} \xi_\beta \end{pmatrix}, \\
w_{ip} =& \frac{h_{M,i} h_{t,i}}{|B_i|} \omega_\alpha \omega_\beta,
\end{eqnarray}
where $p = (\alpha,\beta)$ indexes the $4^2 = 16$ quadrature nodes, $\bar{M}_i, \bar{t}_i$ are the bin midpoints, $h_{M,i} = (M_{\star,\mathrm{hi}}^{(i)} - M_{\star,\mathrm{lo}}^{(i)})/2$ and $h_{t,i} = (t_{\mathrm{hi}}^{(i)} - t_{\mathrm{lo}}^{(i)})/2$ are the half-widths, and ${\xi_\alpha, \omega_\alpha}$ are the standard Gauss--Legendre nodes and weights on $[-1,1]$. The bin-integrated kernel matrix entry is then
\begin{equation}
\tilde{k}(B_i, B_j) = \sum_{p=1}^{16} \sum_{r=1}^{16} w_{ip} ~ w_{jr} ~ k(\mathbf{x}_{ip} \mathbf{x}_{jr}).
\end{equation}

Hyperparameter gradients are computed analytically by applying the same quadrature weights to the gradient tensor of the base kernel, enabling exact (within quadrature error) gradient-based optimization of the log marginal likelihood without numerical differentiation.

The base kernel $k$ is the squared-exponential RBF kernel with independent length scales $(\ell_M, \ell_t)$ for the mass and time dimensions. Both inputs are standardized to zero mean and unit variance prior to fitting, and the GP is trained with \texttt{normalize\_y=True} to handle amplitude variation without an additional constant kernel (which would otherwise double-count the amplitude degree of freedom). Heteroscedastic observational noise is incorporated via the diagonal noise term $\alpha_i = \sigma_i^2$, where $\sigma_i$ is the Poisson uncertainty on the \gls{MI} fraction in bin $i$. Hyperparameters are optimized by maximizing the log marginal likelihood with 15 random restarts from $\ell \in [0.001, 10]$ in standardized units. After training, the GP is evaluated at arbitrary $(M_\star, t)$ points using zero-width bin entries (i.e., point evaluations of the latent field), yielding a posterior mean $\mu(M_\star, t)$ and standard deviation $\sigma_{\rm GP}(M_\star, t)$.


\bibliography{statement,references,references_2}{}
\bibliographystyle{aasjournalv7}



\end{document}